# Atmosphere expansion and mass loss of close-orbit giant exoplanets heated by stellar XUV:

## II. Effects of planetary magnetic field; Structuring of inner magnetosphere

M. L. Khodachenko[1,2], I. F. Shaikhislamov[3], H. Lammer[1], P. A. Prokopov[3]

1) Space Research Institute, Austrian Acad. Sci., Graz, Austria
2) Skobeltsyn Institute of Nuclear Physics, Moscow State University, Moscow, Russia
3) Institute of Laser Physics SB RAS, Novosibirsk, Russia

E-mail address: maxim.khodachenko@oeaw.ac.at

**Abstract:**
This is the second paper in a series where we build a self-consistent model to simulate the mass loss process of a close orbit magnetized giant exoplanet, so-called Hot Jupiter (HJ). In this paper we generalize the hydrodynamic (HD) model of a HJ's expanding hydrogen atmosphere, proposed in the first paper (Shaikhislamov et al. 2014), to include the effects of intrinsic planetary magnetic field. The proposed self-consistent axisymmetric 2D MHD model incorporates radiative heating and ionization of the atmospheric gas, basic hydrogen chemistry for the appropriate account of major species comprising HJ's upper atmosphere and related radiative energy deposition, as well as $H_3^+$ and Lyα cooling processes. The model also takes into account a realistic solar-type XUV spectrum for calculation of intensity and column density distribution of the radiative energy input, as well as gravitational and rotational forces acting in a tidally locked planet-star system. An interaction between the expanding atmospheric plasma and an intrinsic planetary magnetic dipole field leads to the formation of a current-carrying magnetodisk which plays an important role for topology and scaling of the planetary magnetosphere. A cyclic character of the magnetodisk behavior, comprised of consequent phases of the disk formation followed by the magnetic reconnection with the ejection of a ring-type plasmoid, has been discovered and investigated. We found that the mass loss rate of an analog of HD209458b planet is weakly affected by the equatorial surface field <0.3 G, but is suppressed by an order of magnitude at the field of 1 G.

## 1. Introduction

The constantly growing number of discovered exoplanets and accumulation of data regarding their physical and orbital characteristics provide an empirical background for more detailed investigation of general principles and major trends in formation and evolution of the planetary systems, including the potential habitability aspect of the terrestrial type planets. More than a half of known exoplanets have orbits around their host stars shorter than 0.6 AU. By this, an evident maximum in the orbital distribution of exoplanets takes place in the vicinity of 0.05 AU, with two well pronounced major populations there with orbital periods $P < 30$ days corresponding to the giant type planets ($0.2 M_J < M_p < 8 M_J$), so called "Hot Jupiters" (HJs), and less massive ($0.008 M_J < M_p < 0.08 M_J$) Neptune- and Super-Earth type planets. Here $M_J$ stays for the mass of Jupiter. Altogether the HJs comprise about 16% of the total number of known exoplanets.

Close location of the majority of known HJs to their host stars and caused due to that intensive heating, ionization and chemical modification of their upper atmospheres by the stellar X-ray/EUV (XUV) radiation, leads to the hydrodynamic expansion of the ionized atmospheric material. It results in a supersonic planetary wind, which contributes to the so-called *thermal* escape of the planetary atmosphere and related mass loss. Above the exobase and the magnetopause boundary (if it exists), i.e. in the region of direct interaction of the expanding atmosphere with the stellar wind, the escaping particles are picked up by the stellar wind plasma flow resulting in a *non-thermal* mass loss. The planetary wind has been studied by means of hydrodynamic (HD) models in Yelle (2004), Tian et al. (2005), García Muñoz (2007), Erkaev et al. (2005, 2007), Penz et al. (2008), Murray-Clay et al. (2009), Guo (2011, 2013), Koskinen et al. (2010, 2013), as well as in the first paper of this series, Shaikhislamov et al. (2014). All the proposed HD models indicate that hydrogen atmospheres of close-in giant exoplanets are in the regime of continuous expansion and outflow with the corresponding *thermal* mass loss rates $\sim 10^{10} - 10^{11}$ g s$^{-1}$. At such rates a Jupiter mass planet loses about 0.16 % of its mass per Gyr, which is far below the sensitivity of the modern measurement methods. At the same time, the *thermal* escape by exposing the atmospheric material to the oncoming stellar wind appears an important factor and prerequisite for the *non-thermal* mass loss. Therefore, both effects need to be properly considered in their mutual relation during estimation of the total planetary mass loss.

The intrinsic magnetic field is another crucial factor which influences the mass loss of a close-orbit HJ and protects the planetary upper atmospheric environment. The protective role of planetary magnetic field has two major aspects. First, the large-scale magnetic fields and electric currents, related with the planetary magnetism, form the planetary magnetosphere which acts as *a barrier* for the upcoming stellar wind. Magnetosphere protects the ionosphere and upper atmosphere of a planet against direct impact of stellar plasmas and energetic particles, thus reducing the *non-thermal* mass loss (Lammer et al. 2007, Khodachenko et al. 2007a, Khodachenko et al. 2007b). Second, the internal magnetic field of a HJ's magnetosphere strongly affects the *thermal* mass loss, influencing the *streaming of the expanding planetary wind plasma* and its interaction with the stellar wind.

As it was suggested by Khodachenko et al. 2012, the interaction of outflowing partially ionized atmospheric plasma of a HJ with the planetary magnetic dipole field leads to the development of a current-carrying magnetodisk. The inner edge of magnetodisk is located at the so called Alfvénic surface ($r = R_A$) where the kinetic energy density of the moving plasma becomes equal to the energy density of the planetary magnetic field. Two major regions with different topology of the magnetic field (Mestel, 1968) can be distinguished in the magnetosphere of a HJ, driven by the escaping plasma flow. In the so-called "dead-zone", the magnetic force is strong enough to lock plasma with the planet and to keep the field lines closed. In the "wind-zone", the expanding plasma drags and opens the magnetic field lines, leading to the formation of a thin current-carrying magnetodisk. The field of magnetodisk can exceed significantly the dipole field. Altogether, this leads to the development of a new type of *a wind-driven, magnetodisk-dominated* magnetosphere of a HJ, which has no analogues among the solar system planets (Khodachenko et al., 2012). Such expanded magnetodisk-dominated magnetospheres of HJs, which may be up to 40-70% larger, as compared to the traditional dipole-type ones (Griemeier et al., 2004; Khodachenko et al., 2007a), are expected to provide better protection of close-orbit HJs against of the erosive action of the stellar winds (Khodachenko et al., 2012) and related *non-thermal* mass loss.

Two basic processes, acting simultaneously, are responsible for the formation of HJ's magnetodisk: 1) the thermal expansion and escape of the planetary plasma, heated by the stellar radiation, and 2) the centrifugal acceleration of plasma in the co-rotation region by the rotating planetary magnetic field, with the subsequent release of the material beyond the centrifugal Alfvénic surface (the so called "sling" mechanism) (Khodachenko et al., 2012). A simultaneous self-consistent description of both mechanisms is a complex physical problem. So far, these processes were treated separately.

It is especially possible in the case of close-orbit tidally locked planets, which are the subject to strong radiative energy deposition, whereas the rotational effects are usually much weaker, as the planetary rotation is synchronized with the orbital revolution (Khodachenko et al. 2012, Antonov et al. 2013, Shaikhislamov et al. 2014). In this case, the radial expansion of hot planetary plasma dominates the co-rotation in the inner magnetosphere. Besides of a semi-analytical treatment of the inner (dipole-dominated) and outer (magnetodisk-dominated) parts of the HJ's magnetosphere (Khodachenko et al. 2012), the formation of magnetodisk under the conditions of an expanding plasma flow in a background magnetic dipole field was demonstrated for the first time in the laboratory experiment, performed with the relevant for a HJ value of $R_A$ equal to 3 planetary radii, $R_p$ (Antonov et al. 2013).

One of the main goals of the present work is to develop a self-consistent numerical model of an expanding HJ's plasma wind and its interaction with the intrinsic planetary magnetic field. The proposed so far different modeling approaches and approximations for the simulation of expanding planetary atmosphere and related *thermal* mass loss have already addressed various important factors, such as atmospheric composition and photo-chemistry, boundary conditions, spectrum of ionizing radiation and distribution of primary energy input, radiative and adiabatic cooling processes, as well as tidal force. A number of self-consistent models have been developed (Yelle 2004, García Muñoz 2007, Koskinen et al. 2010, Shaikhislamov et al 2014) which doesn't rely anymore on artificial simplifications and idealizations such as mono-energetic XUV flux, isothermal flow, and specific boundary conditions to initiate the wind. However, another crucial factor – the intrinsic planetary magnetic field – usually was not considered at all, or was included in a non-self-consistent way (e.g., by prescribing of specific magnetic configurations). At the same time, the importance of an appropriate account of the planetary magnetism and its role in the mass loss models becomes nowadays widely recognized. To include magnetic field, one has to employ instead of the widely used 1D spherical geometry of the problem and non-magnetized hydrodynamics (HD), at least an axisymmetric 2D geometry, and to treat the problem within the frame of magnetohydrodynamics (MHD).

In the first attempts to incorporate planetary magnetic field a specific topology of the HJ's inner magnetosphere with predefined "wind-" and "dead-zones" was assumed, and the mass loss of planetary atmosphere was investigated in a semi-analytic way (Trammell et al. 2011). Adams (2011) considered outflows from close-in gas giants in the regime, where the flow is controlled by a static magnetic field. The isothermal Parker solution was constructed along the open magnetic field lines, which gives a faster super-sonic transition (Adams 2011), than in the spherical expansion case. However, as it was shown by Trammell et al. (2011), the tidal force can significantly decelerate the polar wind. In agreement with the previous theoretical estimates, it was shown that for sufficiently strong fields the material outflow is confined to open field lines, whereas the inner equatorial region contains a "dead-zone" made of closed field lines and filled with practically not-moving plasma. That results in significantly reduced mass loss rate, as compared to a pure HD case. Estimation of the size of the "dead-zone" and construction of the solution in the "wind-zone" made it possible to estimate the overall mass loss as a function of planetary magnetic field strength (Trammell et al. 2011).

More self-consistent treatment based on 2D MHD codes has been recently performed by (Trammell et al. 2014, Owen and Adams 2014) in which "dead-" and "wind-zones" have been shown to form in the expanding planetary wind. However, the thermosphere heating and the hydrodynamic flow initiated close to the planetary surface were simulated with rather simplified models, assuming a mono-energetic XUV flux, homogenous (e.g., Trammell et al. 2014), or empirically estimated gas temperature, and variable boundary conditions at the planet surface. Note, that the last are known to influence the expanding planetary wind solution (Adams 2011, Trammell et al. 2011, Shaikhislamov et al. 2014). As a result, the obtained estimations for the magnetic field, at which the planetary wind of HD209458b is significantly suppressed, varies in different papers by more than an order of magnitude – from less than 0.3 G in Owen and Adams (2014) up to more than 3 G in Trammell et al. (2014). At the same time, the major conclusion of these more advanced simulations stays the same – planetary *thermal* mass loss can be significantly suppressed by the presence of magnetic field relative to the quasi-spherical HD outflows that are commonly used in modelling planetary evaporation (e.g. Koskinen et al. 2007, García Muñoz (2007), Murray-Clay et al. 2009, Owen & Jackson 2012, Shaikhislamov et al. 2014). Despite of the recognition of importance of the "dead-" and "wind-zones" in the context of the planetary atmospheric mass loss and magnetosphere topology, another important structure – magnetodisk – which is closely associated with these regions, has not been investigated so far.

It should be emphasized, that the processes of material escape and planetary magnetosphere formation have to be considered jointly in a self-consistent way in their mutual relation and influence. In that respect, the self-consistent modelling of HJ's planetary wind, driven by the stellar radiative heating in the presence of an intrinsic magnetic field still remains an actual task. It is a goal of our *two-step modelling* approach, starting with a 1D HD model developed in (Shaikhislamov et al. 2014), which is generalized here to the case of a magnetized planet. The proposed 2D MHD model includes the basic hydrogen photo-chemistry for an appropriate account of atmospheric composition and $H_3^+$ cooling; distributed primary energy input in the system by a realistic Sun-based spectrum of XUV radiation; stellar-planetary gravitational and rotational forces; as well as such boundary conditions, that doesn't influence solution (Shaikhislamov et al. 2014). There is also an external and so far unaccounted factor of the stellar wind plasma flow and CMEs which force the magnetosphere around a planet and control the *non-thermal* mass loss. However, even without this element, the process of planetary wind formation and physical conditions in the immediate vicinity of a HJ, themselves, require a special study, which constitutes a necessary step for better understanding of exoplanetary magnetospheres. While the basic effects of a background planetary magnetic dipole field, such as formation of an inner "dead-zone", which contains a stagnated not-moving plasma, or buildup of an equatorial magnetodisk, can be understood qualitatively using simple physical reasoning and estimates (Mestel 1968, Adams 2011, Trammell et al, 2011, Khodachenko et al. 2012), or by means of laboratory experiment (Antonov et al. 2013), still only a numerical simulation can provide the reliable quantitative results on that matter. In the present paper we report on such a modelling

For the sake of definiteness, we take as typical physical parameters of a modelled HJ those of HD 209458b orbiting a Sun-like G-star. As it will be shown below, the radiative heating and reabsorption, as well as the adiabatic and $H_3^+$ cooling effects included in the model, play certain role in the formation and structuring of the planetary plasma wind. Simulation of tidally-locked analogs of the HD 209458b without magnetic field reveals the development of a strong zonal flow which effectively redistributes the absorbed radiation energy towards the night side. The intrinsic planetary magnetic field is known to influence the expanding wind flow. That has been already studied by Owen and Adams (2014). In particular, the field suppresses the zonal flow, so that the heat, radiatively deposed on the day side, is not efficiently transported to the night side (Owen and Adams 2014), and the material outflow from the night side is thereby reduced. Besides of that, the specific topology of the inner magnetosphere with a "dead-zone" filled with stagnated plasma, and the material outflow, taking place only along the open field-lines of the limited "wind-zone", as well as the effect of the adiabatic cooling due to higher divergence of the dipole-type magnetic field guiding the planetary wind in the polar regions (Shaikhislamov et al. 2014), result in an overall decrease of the atmospheric material escape rate. In our simulations, an intrinsic dipole magnetic field of 1 G on the planetary surface at the equator (i.e. at $r = R_p$) reduces the overall mass loss rate by about an order of magnitude as compared to the non-magnetized case. Our simulations also demonstrate that a thin equatorial magnetodisk is a persistent structure which forms beyond the "dead-zone". The flux captured in magnetodisk reaches up to 10% of the intrinsic dipole field flux across the polar hemisphere of the planet. An essentially novel feature found in the present work, is that a HJ's magnetodisk appears to be a dynamic structure, which undergoes periodic explosive reconnection events accompanied by the release of ring-type current-carrying plasmoids. To certain extend, this is similar to the phenomenon of reconnection in the planetary magnetotail. Therefore, in the present work we paid special attention to the investigation of the dynamical regimes of the HJ's magnetodisk.

The paper is organized in the following way. In Section 2 we describe the modelling concept and address the considered geometry approximations of the proposed 2D MHD model, as well as the included hydrogen chemistry and quasi-realistic stellar XUV radiation flux. In Section 3 the results of numerical simulations are presented and the cases of non-magnetized and magnetized planetary wind are compared. Section 4 is devoted to the discussion of the modelling results and conclusions.

## 2. Modelling concept

To investigate the structure of a HJ's inner magnetosphere formed as a result of interaction and mutual influence of the expanding planetary plasma wind and intrinsic magnetic dipole field, we develop an MHD model as an extension of the HD model, proposed in Shaikhislamov et al. (2014). Besides of the effects of the intrinsic planetary magnetic field, the proposed MHD model includes a realistic spectrum of the stellar XUV radiation (to calculate correctly the intensity and column density distribution of radiative energy input), basic hydrogen chemistry (for the appropriate account of atmosphere composition), $H_3^+$ cooling, as well as stellar-planetary tidal and rotational forces.

The model equations are the same as those solved in Shaikhislamov et al. (2014), supplemented by the equations for the hydrogen chemistry and the equation for the magnetic field. The extended momentum and energy equations (Eq. (3) and Eq.(4) in Shaikhislamov et al. (2014)) include now the Ampere force and more general radiative heating terms (specified below), respectively. To calculate the poloidal magnetic field and azimuthal current in the planet-based cylindrical coordinate system $(r,\varphi,z)$ with $z$-axis co-directed with the magnetic dipole moment, an axisymmetric vector potential $A_\varphi \mathbf{e}_\varphi$ is propagated in time:

$$\left(\frac{\partial}{\partial t} + V_z \frac{\partial}{\partial z} + V_r \frac{\partial}{r \partial r} r\right) A_\varphi = \frac{c^2}{4\pi \sigma} \Delta A_\varphi \qquad (1)$$

By definition, $\mathbf{B} = \mathrm{rot}(\mathbf{e}_\varphi A_\varphi)$ ensures automatic flux conservation, though at the expense of stronger stability requirements. Because of the extremely large system size, the magnetic Reynolds number, Re, of the problem is everywhere exceedingly large, except of a region very close to the planet where the particle density sharply rises. Thus, the dynamics of the magnetic field should be dissipation-less which in numerical model is achieved by taking sufficiently high, though finite, conductivity which corresponds to $\mathrm{Re} \geq 10^5$. Close to the planet surface where electron-atom collisions dominate as compared to the electron-ion collisions, we retain the electric conductivity $\sigma$ calculated with the corresponding expressions from Braginskii (1965). An open boundary condition is applied at the outer boundary of the simulation domain. In order to avoid its numerical influence on the simulation result (in case with magnetic field), we place the outer boundary at the empirically determined (large enough) distance of at least 20 $R_\mathrm{p}$. The applied numerical solver of our simulation code treats MHD equations discretized on a uniform cylinder mesh $r$, $Z$ with a step of $R_\mathrm{p}/200$ or smaller. Such small mesh is required to resolve XUV energy absorption regions in the dense highly stratified atmosphere close to the conventional surface of a planet. An explicit numerical scheme is employed with an upwind calculation of fluid dynamics and a prediction sub-step for magnetic values. The robust stability was achieved by restricting the time step by ¼ of the Courant limit.

To study the influence of magnetic field on the expansion of atmosphere and mass loss of a close-orbit giant exoplanet, we take for definiteness sake, as a typical example, the planet HD 209458b with the mass of $M_p = 0.71 M_J$ and radius of $R_p = 1.38 R_J$ which orbits its star at 0.047 AU. The parent star of this planet, HD 209458, is a solar-type G-star with the mass of $M_{st} = 1.148 M_{Sun}$ and the age of ~4 Gyr (The Extrasolar Planets Encyclopaedia (2014): http://www.exoplanet.eu). The major physical parameters and results of the simulations further on will be scaled in units of the characteristic values of the problem, defined as follows: for temperature, $\tilde{T}_c = 10^4$ K; for speed, $V_{Ti}(\tilde{T}_c) = \tilde{V}_c = 9.1$ km/s ; for distance, the planet radius $R_p$; for time, the typical flow time $\tilde{t}_c = R_p / \tilde{V}_c \approx 3$ hr.

### 2.1 *Geometry assumptions of the modelled case and account of gravitational effects*

Inclusion of magnetic field in the model of an expanding upper atmosphere of a HJ, driven by stellar XUV, requires changing from the commonly used and relatively simple for the numerical algorithms 1D spherical geometry to the more sophisticated, 2D or 2.5D cylinder or fully 3D ones. In fact, the problem of an expanding magnetized atmosphere of a rotating HJ is essentially 3D one, and it is not possible to find an appropriate mutual orientation of the planet's rotation, orbiting, and magnetic dipole axes, for which at least an axisymmetric 2.5D approach could be fully valid. Moreover, the widely used assumption of a spherically symmetric homogenous XUV energy deposition appears too strong simplification for the tidally locked systems where a planet is illuminated by star from just one side (day-side). In the most common case of planetary magnetic dipole and rotation axes oriented perpendicular to the ecliptic plane, the stellar radiation flux will have in the planet-based reference frame either a continuous day-night dichotomy or diurnal variations. In a more specific, though not forbidden, case of a tidally locked planet with the magnetic dipole moment oriented along the planet-star line the 2D symmetry is destroyed because of the Coriolis force which appears in the planet-based reference frame due to the rotation of planet synchronous with its orbital revolution.

In the present study we employ a 2D axisymmetric MHD code for the treatment of a tidally locked close-orbit exoplanet in the following two cases of the problem's geometry approximations:

*1)* In the first case the symmetry axis of the numerical problem is taken along the planet-star line in the reference frame of the rotating (synchronous with the orbital revolution) tidally locked planet (see Figure 1a). Within this approximation we disregard the Coriolis force and average the centrifugal force as shown in Figure 1a (in the terminator plane). Such simplification is possible for sufficiently slowly rotating planets. In particular, the surface rotation velocity varying along the terminator line, which appears a symmetry breaking factor, is usually much less than the velocities scale in the system. For example, for HD209458b it is ~ 3 km/s (on equator); that is less than the typical outflow velocity of ~10 km/s. In that respect, the Coriolis force is not expected to play significant role. Moreover, it doesn't change the energy of the moving material, so its influence on the thermal mass loss rate is negligible. Another symmetry-breaking factor for a slow rotating tidally locked HJ is the centrifugal force. It has a maximum value in the ecliptic plane and turns to zero on the rotation axis. The applied circular averaging around the planet-star line (as indicated in Figure 1a) disregards this difference, which is also not crucial for the considered processes of mass loss. Further on we will call such approach as a *quasi-axisymmetric* approximation. It is especially suited for the study of mass loss of tidally locked non-magnetized planets, illuminated by the star from only one fixed side (day-side). It can be applied also for the magnetized planets with magnetic moment oriented along the planet-star line, but not to the planets with that perpendicular to the ecliptic plane.

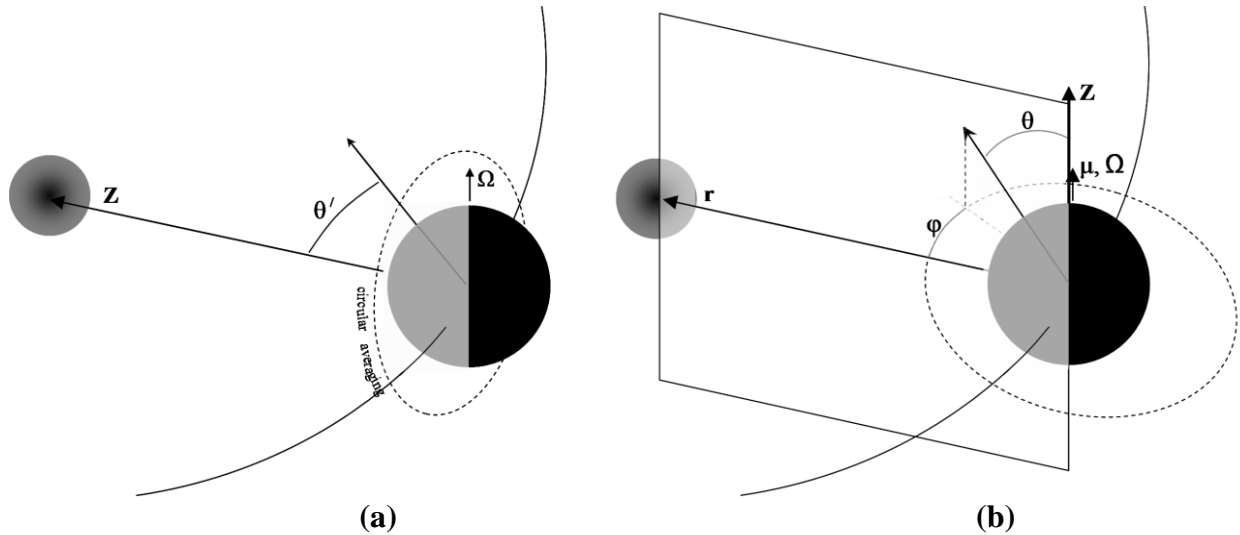

**Figure. 1** Two cases of geometry approximation applied for calculation of planetary wind and related mass loss: *(a) quasi-axisymmetric* approximation, suitable for a non-magnetized planet or a planet with magnetic dipole oriented along the planet-star line; and *(b) sector* approximation suitable for a magnetized planet.

2) In the second case the symmetry axis of the numerical problem is directed perpendicular to the ecliptic plane (see Figure 1b). We use this geometry to study magnetized planets. The simulation results in this case are restricted by the day-side part of the meridianal plane (i.e. semiplane $X(>0)$-$Z$ in a supplementary planet-based Cartesian system with $X$ directed towards the star), assuming that no variation of physical quantities and no material motion in azimuthal, i.e. across the meridianal plane, or in $\varphi$-direction in Figure 1b, take place. Further on we will call such approach as a *sector* approximation. Similar geometry consideration was realized in Erkaev et al. (2013) and Owen and Adams (2014), but it is different from the approach of Trammell et al. (2014), where boundary conditions at the planetary surface and tidal force were assumed to be rotationally averaged. Since in the present paper we investigate the tidally locked HJs, no rotational averaging can be (strictly) applied. Sector approximation also does not allow inclusion of the effect of plasma co-rotation with the planetary magnetic field, which in any case remains negligibly small for the slowly rotating tidally locked planet. The applicability of sector approximation for the magnetized planets can be partially justified by the fact, that sufficiently strong planetary dipole magnetic field suppresses significantly the azimuthal flows which otherwise develop in the system. It will be shown in sub-section 3.1.2, that in the non-magnetized case the calculations done with the *sector* geometry approximation reproduce well the results obtained with a more adequate *quasi-axisymmetric* approximation, which includes also the radiatively unheated night-side hemisphere and the global material motion. In particular, the over-estimating effect of the *sector* approximation regarding the mass loss is not very strong, because the incompletely reproduced zonal flow between the day- and night-sides, while effectively redistributing the heated material (or in other words, the heating), does not affect much the overall material expansion and the related mass loss. Therefore, the proposed modelling of the planetary wind in the presence of magnetic field is useful, as well as the similar one, realized in Owen and Adams (2014), and in Trammell et al. (2014). The advantage of the study, reported in the present paper which was aimed to investigate first of all the magnetic field effects, consists in avoiding of any quasi-empirical specific assumptions regarding gas dynamics, magnetic field topology, and boundary conditions.

The description of gravitational stellar-planetary interaction, as well as of the planetary rotation and orbital motion effects, are directly related to the applied geometry approximations. For a tidally-locked close star-planet system this can be done by means of the generalized potential included in the momentum equation (Eq.(3) in Shaikhislamov et al. (2014)) and expressed in the planet-based spherical coordinate system $(R,\theta,\varphi)$, in which the host star is located at $\theta = \pi/2$, $\varphi = 0$, $R = D$, where $D$ is the planetary circular orbit distance. The first order approximation of the generalized potential over small parameter $R/D$ yields:

$$U = -\frac{GM_p}{R} \cdot \left(1 + \frac{1}{2}\frac{R^3}{R_{L1}^3}\frac{(1+3\cos^2\varphi)\cdot\sin^2\theta - 1}{3}\right), \qquad (2)$$

where $R_{L1} = D(M_p/3M_*)^{1/3}$ is the first Lagrange point. Further on, we will call the part of the generalized potential which contains $R_{L1}$ as the tidal force. It has to be additionally specified for the different geometry approximation cases, defined above and used in the modeling.

For the day-side *sector* geometry approximation, the gravitational potential defined by Equation (2), in the planet-based cylindrical coordinate system with $r = R\sin\theta$ and Z-axis directed perpendicular to the ecliptic plane (see Figure 1b), has to be taken at a particular azimuth angle $\varphi = 0$. In the *quasi-axisymmetric* geometry approximation with Z-axis of the planet-based cylindrical coordinate system directed along the planet-star line, Equation (2) has to be averaged around it and expressed in terms of the corresponding polar angle $\cos^2\theta' = \cos^2\varphi\cdot\sin^2\theta$. That which gives

$$U = -\frac{GM_p}{R} \cdot \left(1 + \frac{1}{2}\frac{R^3}{R_{L1}^3}\frac{7\cos^2\theta' - 1}{6}\right) \qquad (3)$$

Whenever necessary, for presenting results we will use either the coordinates $(R,\theta)$ of the spherical or $(r,Z)$ of the cylindrical frame with $R$ and $r$ denoting spherical and cylindrical radii, respectively.

**2.2 Radiative (XUV) energy deposition**

The XUV radiation of a host star affects the energy budget of the planetary upper atmosphere, resulting in the mutually dependent and interrelated 1) heating, 2) ionization and 3) expansion of the atmospheric gas (Yelle 2004, Erkaev et al. 2013, Tian et al. 2005, García Muñoz 2007, Penz et al. 2008, Guo 2011, 2013, Koskinen et al. 2010, 2013, Lammer et al. 2013, Shaikhislamov et al. 2014). By this, the location and size of the energy deposition region where the radiative energy is absorbed, depends on the local composition of the heated gas (i.e. on the presence of absorbents e.g. atomic or molecular hydrogen), as well as on the spectral (energetic) features of the penetrating photons. The gas ionization degree has to be properly calculated from ionization/recombination and advection processes, because it not only controls the XUV absorption and heating, but also influences the total gas pressure by affecting the additional electron pressure gradient in the ionization regions. Basic issues regarding this effect and the mechanism of radiative heating have been investigated and summarized in Shaikhislamov et al. (2014). It has been shown, that to completely take into account the radiative energy deposition, the inner boundary of the simulation domain (i.e. conventional planet surface) has to be defined at the height, which is not reachable for the modeled XUV radiation.

To make our model more reliable for the real stellar/planetary conditions and to avoid of any empirical *ad hoc* substitutions and approximations, so that the intensity and column density distribution of primary energy input in the system could be calculated directly, we include in the proposed MHD numerical model the XUV heating term based on the realistic spectrum of the stellar radiation. As a proxy for the ionizing EUV flux produced by a sun-like star we use a compilation spectrum by (Tobiska 1993) which covers 10–912 Å range, binned by 1 Å. It is based on measurements of the solar radiation under moderate activity conditions with the proxy index $P_{10.7} = 148$. The total integrated flux at 1 AU in this model is 4.466 erg s$^{-1}$ cm$^{-2}$; that is close enough to the time averaged total solar flux (3.9 erg s$^{-1}$ cm$^{-2}$) shortward of 912 Å (Ribas et al 2005).

In the present study we suppose for a HJ a pure hydrogen atmosphere. Absorption by hydrogen atom with a differential cross-section $\sigma = \sigma_o [E_{ion}/\hbar\nu]^3$, where $\sigma_o = 6.3 \cdot 10^{-18}$ cm$^2$, results in ionization and release of a photo-electron with energy of $E_{ion} - \hbar\nu$ which is shared, with the empirical efficiency of 0.5 (Waite et al. 1983), between all particles assumed to be in thermal equilibrium. The absorption cross-section $\sigma_{H2}$ of the molecular hydrogen $H_2$ has been taken to be 2–2.5 times larger than that for H (Yan et al. 1998). The heating term used in the energy balance equation has the following form:

$$Q = \left(n_H + n_{H_2}\, \sigma_{H2}/\sigma_o\right) \cdot q(\sigma_o N_H), \tag{4}$$

where

$$q(\sigma_o N_H) = \int F_{XUV}(\nu) \cdot (\hbar\nu - E_{ion})\sigma_o [E_{ion}/\hbar\nu]^3 \cdot \exp\left(-\sigma_o N_H [E_{ion}/\hbar\nu]^3\right) d\nu,$$

$$\sigma_o N_H = \sigma_o \int \left(n_H + n_{H_2}\, \sigma_{H2}/\sigma_o\right) dl,$$

with $F_{XUV}$, $n_H$, $n_{H2}$ staying for the XUV radiation flux and number densities of atomic and molecular hydrogen, respectively. The photo-ionization rate is split between H and $H_2$ species correspondingly to their cross-sections. Here we ignore the insignificant difference between ionization potential $E_{ion}$ for atomic (13,6 eV) and molecular (15,4 eV) hydrogen. This can be additionally justified by the fact that the $H_2$ dominated layers are reached only by highly energetic photons $\hbar\nu \gg E_{ion}$.

### 2.3 *Hydrogen plasma chemistry*

As an initial state of the simulated atmosphere of a HD209458b analogue planet we take the pure hydrogen gas with the temperature of 1000 K, distributed according to the static barometric equilibrium. The *1 nm* photons from the high-energy edge of the used in our model real solar EUV spectrum are absorbed at a column density of the order of $10^{23}$ cm$^{-2}$ which corresponds to a pressure of 100 μbar. This fact was pointed out in (Shaikhislamov et al. 2014). At the same time, already at the pressures above 10 μbar the simulated atmosphere of a HJ consists of hydrogen molecules. Hence the model has to include the energy and temperature dependent reactions between $H$, $H^+$, $H_2$, $H_2^+$, $H_3^+$, and $e^-$ components, to calculate the composition of the expanding atmosphere of a HJ.

In the present study, we take the particle density at the conventional planet surface (i.e. at the inner boundary of the simulation domain $R=R_p$) fixed at even higher value, $10^{16}$ cm$^{-3}$, which corresponds to the pressure of about 1000 μbar. Such a large density has been chosen to ensure complete absorption of the XUV flux, so that the particular height of the inner boundary location does not affect the solution. With a series of test runs it has been found that 3 times decrease of the boundary pressure changes the calculated mass loss rate only by 2%. Therefore, the inner boundary conditions used in our model are not a variable free parameter.

Exposed to EUV radiation, the $H_2$ atmosphere evolves according to the reactions listed in Table 1, which summarizes the used rates or cross-sections (when relevant) as well as the relative importance of each reaction regarding the mass loss rate: "+" stays for "important", "– " stays for "indifferent" and "–+ " stays for "somewhat affecting".

**Table 1.** Hydrogen chemistry
Rates are given in cgs units. Temperature is scaled in units of $10^4$ K.
"*" corresponds to the cooling rate.
Source: UMIST database: http://udfa.ajmarkwick.net/

| N | Reaction | Product | Rate | Importance | Ref. |
|---|---|---|---|---|---|
| 1 | $H + \hbar\nu$ | $H^+ + e$ | $\sigma_o = 6.3 \cdot 10^{-18}$ | + | |
| 2 | $H_2 + \hbar\nu$ | $H_2^+ + e$ | $\sigma_{H2} \approx (2 \div 2.5)\sigma_H$ | + | Yan et. al. (1998) |
| 3 | $H + e$ | $H + e + \hbar\nu$ | $^*2.9 \cdot 10^{-19} T^{-0.5} e^{-11.84/T}$ | –+ | Dalgarno, McCray (1972) |
| 4 | $H + e$ | $H^+ + 2e$ | $7.3 \cdot 10^{-9} T^{0.5} e^{-15.776/T}$ | – | Voronov (1997) |
| 5 | $H^+ + e$ | $H + \hbar\nu$ | $2.5 \cdot 10^{-13} T^{-0.75}$ | – + | UMIST database |
| 6 | $H_2^+ + e$ | $2H$ | $3.5 \cdot 10^{-9} T^{-0.43}$ | – | UMIST |
| 7 | $H_3^+ + e$ | $H_2 + H$ | $4 \cdot 10^{-9} T^{-0.5}$ | + | UMIST |
| 8 | | $3H$ | $7.6 \cdot 10^{-9} T^{-0.5}$ | + | UMIST |
| 9 | $H_2^+ + H_2$ | $H_3^+ + H$ | $2.1 \cdot 10^{-9}$ | + | UMIST |
| 10 | $H_3^+ + H$ | $H_2^+ + H_2$ | $2.1 \cdot 10^{-9} e^{-1.99/T}$ | – | UMIST |
| 11 | $H_2^+ + H$ | $H^+ + H_2$ | $6.4 \cdot 10^{-10}$ | + | UMIST |
| 12 | $H^+ + H_2$ | $H_2^+ + H$ | $2.1 \cdot 10^{-9} e^{-2.124/T}$ | + | McConnel et. al. (1987) |
| 13 | $H_2 + H$ | $3H$ | $1.4 \cdot 10^{-8} T^{-1} e^{-8.41/T}$ | + | UMIST |
| 14 | $H_2 + H_2$ | $H_2 + 2H$ | $1.4 \cdot 10^{-8} e^{-8.41/T}$ | – | UMIST |
| 15 | $H_2 + 2H$ | $H_2 + H_2$ | $9.5 \cdot 10^{-35} T^{-1.3}$ | – | Tsang & Hampson (1986) |
| 16 | $H^+ + 2H_2$ | $H_3^+ + H_2$ | $3.2 \cdot 10^{-29}$ | – | Kim & Fox (1994) |

Hydrogen composition affects the numerical solution in two ways. First, the deepest layers with densities $>10^{10}$ cm$^{-3}$ located at heights $r \leq 1.1 R_p$ consist predominantly of the molecular hydrogen $H_2$ and have the barometric length scale twice smaller than that in the upper layers, populated by the atomic hydrogen H. At least 10% of the EUV flux at the shortest wavelengths reaches the $H_2$ dominated atmosphere and is absorbed there. This is sufficient to affect the gas motion and the mass loss rate. Second, in accordance with other works (Yelle 2004, García Muñoz 2007, Koskinen et al. 2010), we found that the amount of $H_3^+$ formed in the $H_2$ dominated regions results in significant energy loss due to the infrared cooling – the so-called "$H_3^+$ *cooling effect*". For the cooling rate, we use in our model the result of recent calculation by Miller et al. (2013), who took into account for the first time the non-LTE decrease of total $H_3^+$ radiation depending on the density of $H_2$ and predicted a maximum radiation rate of $5 \cdot 10^{-18}$ W per molecule per radian, reached at the temperature of ~5000 K. This is several times larger than the rate used in works cited above.

Nevertheless, the overall effect of $H_3^+$ cooling on the total thermal mass loss rate for a HD209458b analog exoplanet found in the present study, is about the same as that reported in Yelle 2004, García Muñoz 2007, Koskinen et al. 2010. In particular, the account of $H_3^+$ cooling reduces the mass loss rate by 1.5–2 times, as compared to the case without IR-cooling.

### 3. Simulation Results

#### 3.1 *Non-magnetized HJ's plasma wind*

In order to test the extended 2D model of the planetary expanding wind versus the previously used 1D one and to verify mutual validity of the considered geometry approximations (see section 2.1), as well as to understand the role of various physical factors, such as tidal force and different mechanisms of radiative heating and cooling, we perform the first round of simulations of the HJ's expanding atmosphere without inclusion of the planetary magnetic field (i.e. still the HD modelling).

*3.1.1 Effect of the tidal force*

To understand better the nature and structure of a HJ's environment and its major influencing factors, we performed the 2D HD modelling for the cases with and without the tidal force, i.e. assuming in Equation (3) for the generalized gravitational potential $R_{L1} = 4.53 R_p$ (calculated for HD209458b) and $R_{L1} = \infty$, respectively. The 2D plots for the main physical quantities calculated in the *quasi-axisymmetric* geometry approximation are shown in Figure 2. They demonstrate that the one-side illumination and related heating of the atmosphere by the stellar XUV generates a substantial zonal flow around the planet over the terminator towards the night-side. Due to the adiabatic cooling, compressional heating, and material transport effects, the expanding flow cools the day-side and increases the temperature at the night-side. The zonal wind is so strong, that the temperature and material outflow appear to be higher at the night side of the modelled HJ. The zonal flow produces the large scale vortices at the night-side and a turbulent tail. Because of dense atmosphere and the radiation screening by the planet, the gas flowing tail-wards is only partially ionized (~50%) as compared to the adjacent fully illuminated regions. Partial ionization results in the stronger absorption of the EUV radiation and related additional heating. This can be seen in the temperature and ionization degree plots.

As one can see in Figure 2, the tidal force significantly influences the flow of expanding atmospheric material around an HD209458b type planet. First of all, since the generalized tidal force together with the planet gravity stop the lateral outflow, the planetary wind is blocked sufficiently far from the planet in the direction perpendicular to the planet-star line. This effect was first discussed in Trammell et al. (2011). According to Equation (2), this is true only on the planet rotation axis (i.e. the polar direction; $\theta = 0$) along which centrifugal force is zero. In the direction perpendicular to the planet-star line and planet rotation axis ($\theta = \pi/2, \varphi = \pi/2$) the centrifugal and tidal force effects approximately balance each other, and the generalized tidal force is zero in the first as well as in the second order. Therefore, the escaping wind flow remains confined between two surfaces placed symmetrically relative the ecliptic plane and located at about $16 R_p \approx 4 R_{L1}$ (according to the simulation). Due to the *quasi-axisymmetric* geometry approximation used in the simulations, which assumes an averaging of the tidal force around the planet-star line (see in Figure 1a), these flow-limiting surfaces form a kind of a tube directed towards and outwards the star. Despite the strong modification of the flow by the tidal force, the integral mass loss rate is only about 20% larger and material escape is more or less equally distributed between the day- and night- side directions.

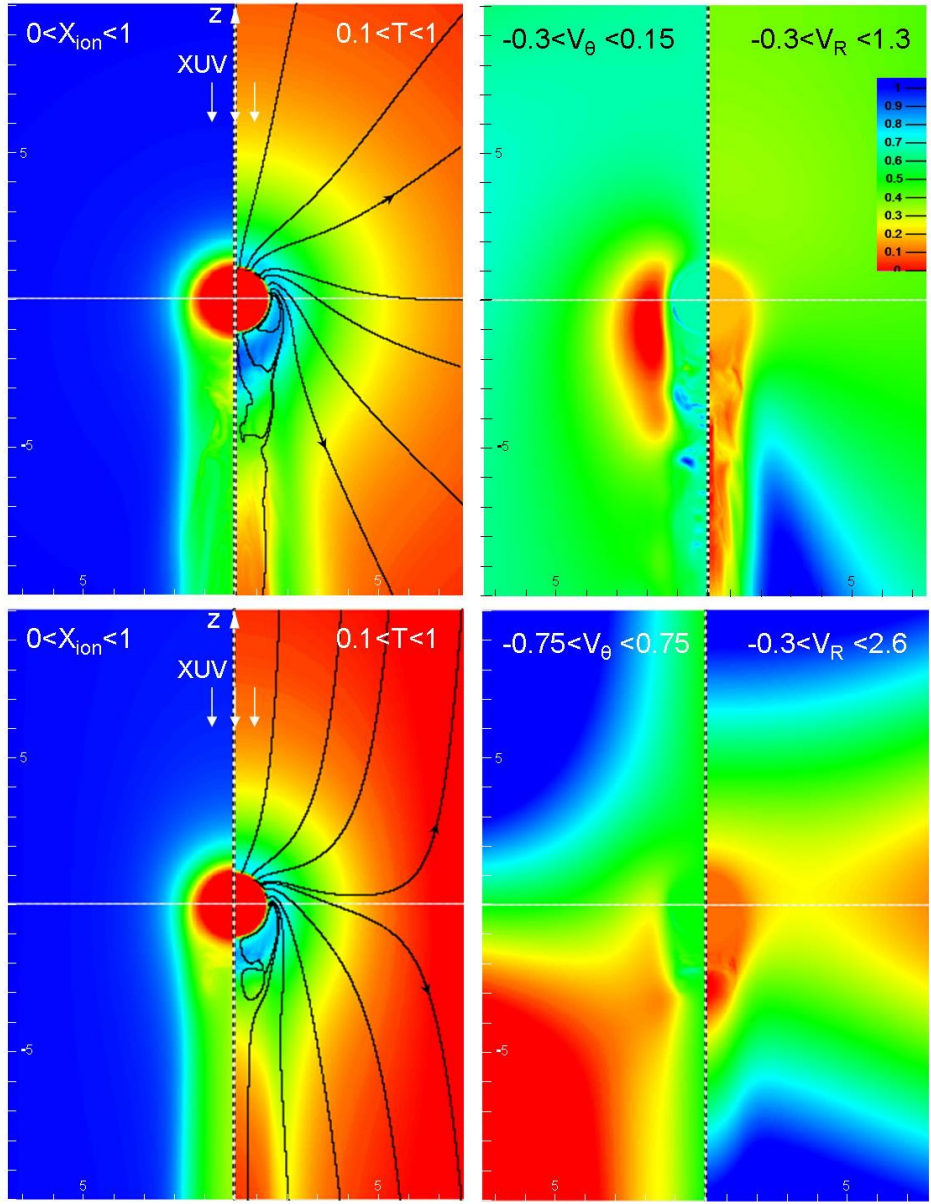

**Figure 2.** Color maps of temperature T, ionization degree $X_{ion} = n_i/(n_i + n_H)$, radial (spherical) Vr and zonal Vθ velocities, calculated in the *quasi-axisymmetric* geometry approximation for a non-magnetized analog of HD209458b in the cases with 1) zero tidal force, i.e. $R_{L1} = \infty$ (top panels), and 2) a realistic tidal force, i.e. $R_{L1} = 4.53 \cdot R_p$ (bottom panels). Variation ranges of the simulated physical parameters are indicated in the top of the corresponding panels. Adopted color scale between the normalized minimal and maximal values of a simulated physical parameter is presented in the upper right panel. All values outside the considered variation range of each parameter, i.e. below or above the corresponding minimal and maximal values, are plotted as uniformly red or blue, respectively. The black lines show the material flow streamlines.

*3.1.2 Sector approximation versus quasi-axisymmetric one: comparison of simulation results*

The escaping non-magnetized planetary wind can be also calculated in the geometry suited for the case with the planetary magnetic field included, i.e. in *sector* approximation. The result of such calculation is presented in Figure 3, which shows spatial distributions of the plasma radial velocity and density. Both demonstrate the same features as those calculated in *quasi-axisymmetric* geometry approximation (see Figure 2). The discontinuity line visible in the Figure 3 is a topological boundary formed due to redirecting of the flow towards the star by the tidal force action, as it was discussed in section 3.1.1. The flow regions outside this boundary are not connected to the planet by streamlines and are populated by several orders of magnitude more rarified background plasma which fills the surrounding space at a fixed minimum level required for numerical processing.

For more detailed quantitative comparison of the simulation results obtained with the *quasi-axisymmetric* and *sector* geometry approximations we plot in Figure 4 the profiles of the HJ's expanding atmosphere density $n$, temperature $T$, and velocity, calculated in the directions towards the star and perpendicular to the planet-star line (i.e. Z-axis). Except of the velocities in the direction perpendicular to the planet-star line, all other parameters of the flow, calculated with both geometry approximations, show good agreement. The difference between the velocities in the direction perpendicular to the planet-star line comes from different treatment of the zonal flow. In the *quasi-axisymmetric* geometry approximation the zonal flow proceeds from dayside to night side across the terminator plane, while in the *sector* geometry approximation the material is assumed to move only from the equator plane to poles, whereas the motion across the meridianal plane is excluded from the consideration.

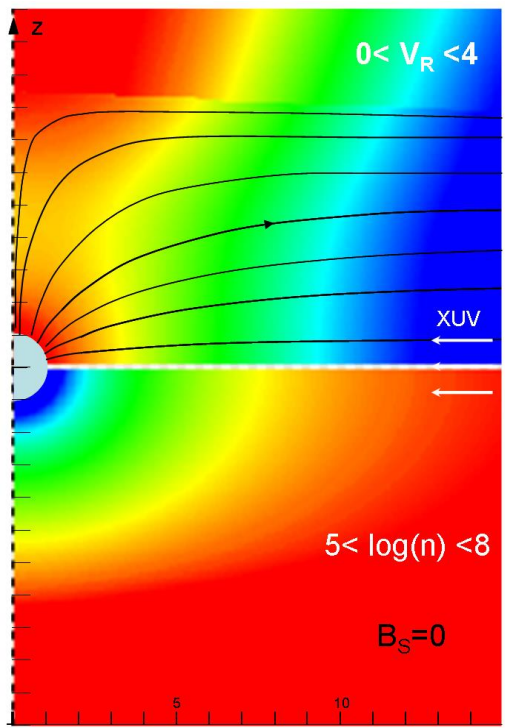

**Figure 3.** Spatial distribution profiles of radial velocity (top panel) and plasma density $n = n_H + n_i$ (bottom panel), calculated for a non-magnetized analog of HD209458b in *sector* approximation geometry with Z-axis directed perpendicular to the ecliptic and the planet-star line. Black lines show the streamlines of the planetary wind flow.

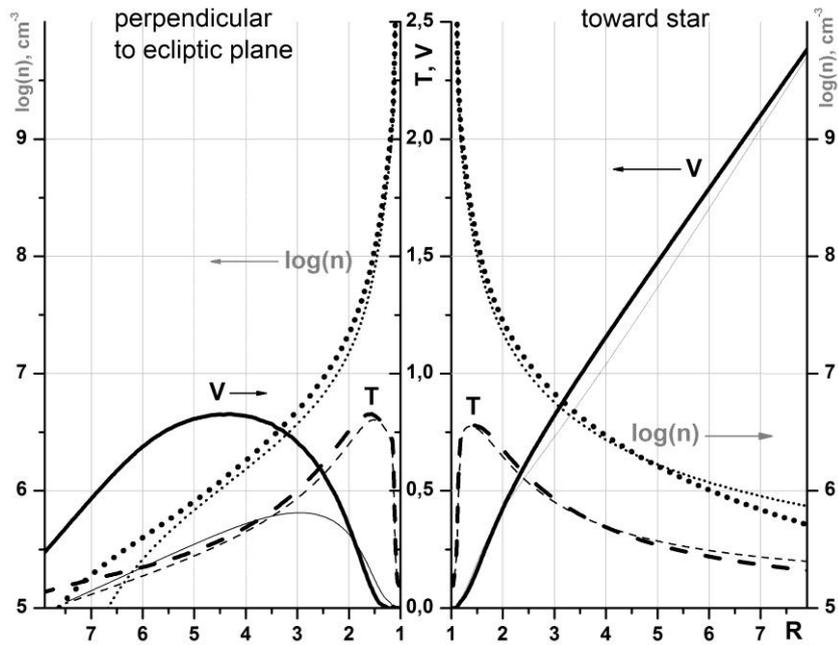

**Figure 4.** Profiles of the planetary wind plasma velocity (solid), temperature (dashed) and density $n = n_H + n_i$ (dotted) in the Z-axis direction, i.e. perpendicular to the planet-star line (left panel) and in the direction towards the star (right panel), scaled in $R_p$. Thick and thin lines correspond to the calculations in *sector* and *quasi-axisymmetric* geometry approximations, respectively. Arrows point to the graph axis which shows scaling of a particular parameter.

There is also a factor of two difference between the tidal force values calculated for the considered different geometry approximations (see Equations (2) and (3)). However, the overall agreement of the calculations performed with two geometry approximations confirms the validity of the *sector* approximation, which will be used for the study of atmospheric material escape in the presence of the planetary magnetic field.

To estimate the total mass loss in the *sector* approximation, we extrapolate the 2D modelling result, obtained for the day-side meridianal plane, to the whole volume over $\varphi$ around the planet, i.e. multiply it by $2\pi$. This approach is similar to that used in 1D modelling, where the total mass loss is determined by multiplying of the 1D solution by $4\pi$. By this, to take into account the reduced material escape on the radiatively unheated night side, a net efficiency of XUV heating is commonly used in the 1D modelling as a kind of correction coefficient. In our simulations the correction coefficient to the overestimated in the case of *sector* approximation mass loss rate (which also neglects the day-night heating difference) could be found from the comparison of the simulation results obtained for both types of the geometry approximation in the non-magnetized case. In particular, the mass loss rate calculated by the above described extrapolation of the *sector* approximation solution to the whole volume appears to be about 2.5 times lager than that obtained under the same conditions with more precise *quasi-axisymmetric* geometry approximation. Therefore, to reflect the effect of the non-uniform illumination and zonal flow in the mass loss rate of a tidally locked exoplanet, calculated by extrapolation of the solution obtained in *sector* geometry approximation, one has to apply the correction coefficient $1 / 2.5 = 0.4$. Note that this correction coefficient is relatively close to an empirical one 0.5, usually applied in other works. Further on we will use the obtained correction coefficient in the calculations for the magnetized case, i.e. with the symmetry axis directed perpendicular to ecliptic plane, and in such a way will roughly take into account the effect of non-uniform illumination of a tidally locked planet and zonal flow.

*3.1.3 Heating and cooling in a self-consistent model*

To demonstrate relative importance of different heating and cooling mechanisms depending on the particular region of the expanding atmosphere, we present in Figure 5 the EUV absorption (Eq.(4)), adiabatic, infrared ($H_3^+$), and Lyα cooling rates as functions of height along the planet-star line, as well as temperature profile and normalized distribution of $H_2$ (relative to the total number density of hydrogen particles $N_{tot}$), calculated under conditions of a non-magnetized analog of HD209458b using *sector* geometry approximation.

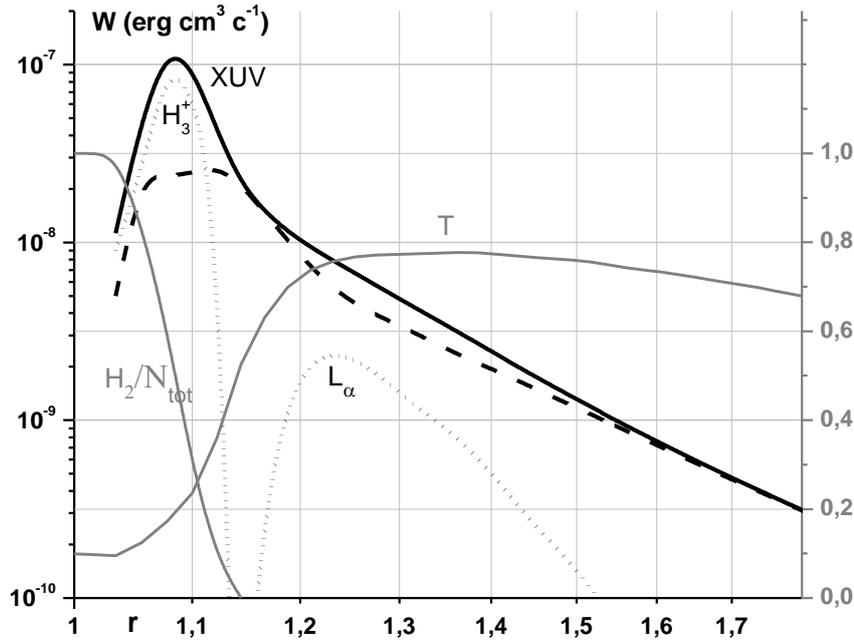

**Figure 5.** Spatial variation of the EUV absorption rate (thick solid line), advection and adiabatic cooling rate (dashed line), as well as radiative cooling rate due to infrared (i.e. $H_3^+$) and Lyα emission (dotted lines) along the planet-star line (scaled in $R_p$), calculated for a non-magnetized analog of HD209458b within *sector* geometry approximation. Gray lines show profiles of temperature and relative content of molecular hydrogen along the planet-star line (values are given at the right axis).

The obtained distributions are close to those found in course of 1D HD modeling in Shaikhislamov et al. (2014). As it can be seen, molecular hydrogen, due to dissociation, is present only at heights below $1.1R_p$, which correspond the pressures above 0.03 μbar. The EUV heating is mostly balanced by advection and adiabatic cooling, though Lyα and especially $H_3^+$ cooling make definite contribution in the certain (not overlapping) regions. The most significant temperature increase takes place exactly between these regions with dominating infrared and Lyα cooling, respectively. Without inclusion of these radiative cooling mechanisms, the maximum temperature value increases by ~1500 K and the position of maximum shifts closer to the planet, while the mass loss rate approximately doubles.

**3.2 *Magnetized HJ's plasma wind***

As a next step, we consider the expanding atmospheric plasma wind in the presence of a planetary magnetic dipole field. To characterize the intrinsic magnetic field of a planet we will use its surface value at equatorial plane $B_s$.

3.2.1 *Magnetic field and plasma flow topology*

In line with qualitative expectations, in the case of a sufficiently small magnetic field, no "dead-zone" is formed. This happens if the gas thermal and ram pressures exceed everywhere the magnetic pressure so, that the last cannot stop the developing flow. For the conditions of an HD209458b analogue planet such situation takes place at $B_s<0.1$ G. The resulting spatial structure of the magnetic field and planetary wind is close to that in the case of radially stretched field lines following the streamlines realized with a broadly distributed azimuthal current $J_\varphi \sim \sin\theta/R^2$ centred on the equatorial plane.

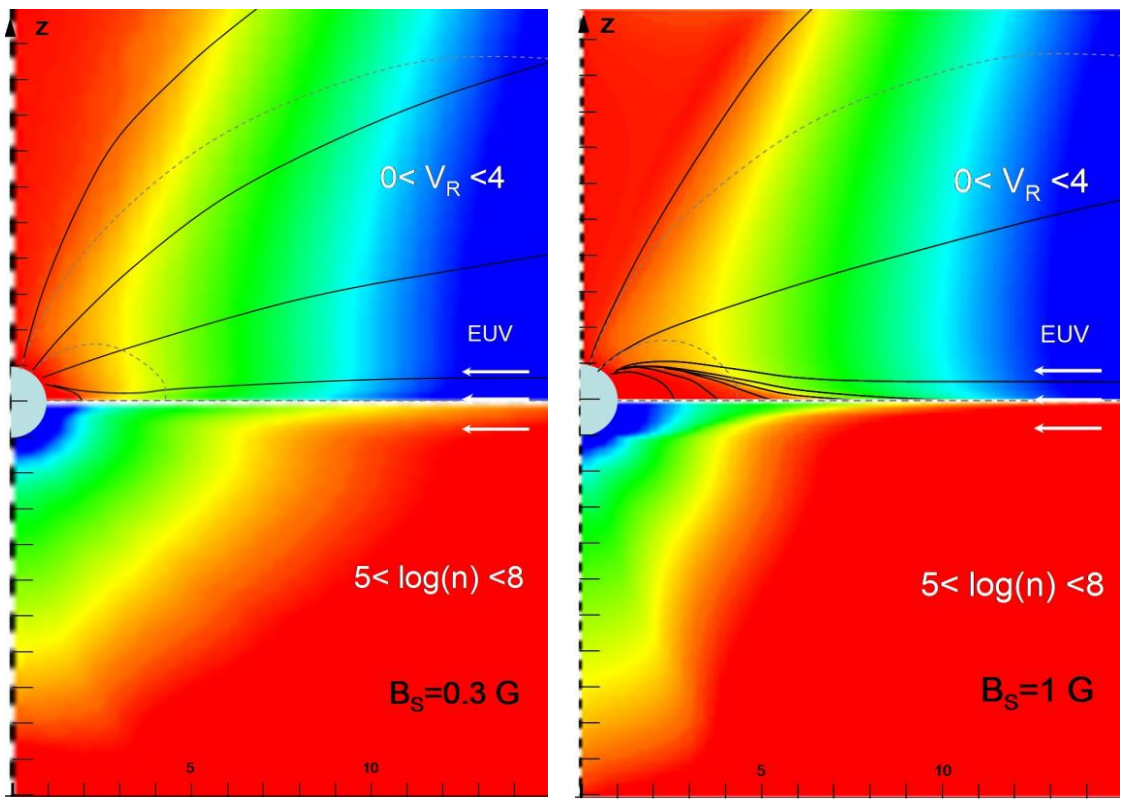

**Figure 6.** Color maps of the radial velocity and particle density distributions calculated for a magnetized analog of HD209458b with equatorial surface field $B_s$=0.3 G (left panel) and $B_s$=1 G (right panel). Magnetic field lines are shown in black. For comparison the dipole field lines are plotted by dashed gray lines.

The left panel of Figure 6 demonstrates the solution obtained with $B_s$=0.3. The planetary wind in this case is moderately influenced by the magnetic field, and a small "dead-zone" is formed. The strongest effect is close to polar regions, where the flow is unable to fully stretch the dipole field lines and near the planet it follows along them. Due to the larger divergence of the material flow controlled by the magnetic dipole field and the corresponding faster adiabatic cooling, as well as because of the suppression of the zonal flow by magnetic force, the bulk velocity in polar regions is significantly diminished and doesn't reach the supersonic values. This effect was predicted in (Shaikhislamov et al. 2014). We note that, if not suppressed by the dipole magnetic field, the polar flow in case of a small tidal force might be stronger than the equatorial wind. Because of the suppression of the polar wind, the gas density above poles is higher than that in the wind zone.

Another difference, as compared to the non-magnetized case, is that due to the rigidity of the open field lines, the planetary wind flow above poles cannot be fully redirected by the tidal force towards the star and it propagates farther away from the planet, even though the velocity is strongly diminished.

At $B_s$=1 G, the "dead-zone", the suppressed polar flow region, and the "wind-zone" between them are well developed and appear to be clearly visible in the right panel of Figure 6. For such value of the magnetic field, the field lines passing near the "dead-zone" are stretched strongly in the vicinity of the equatorial plane and have the shape typical for the current disk.

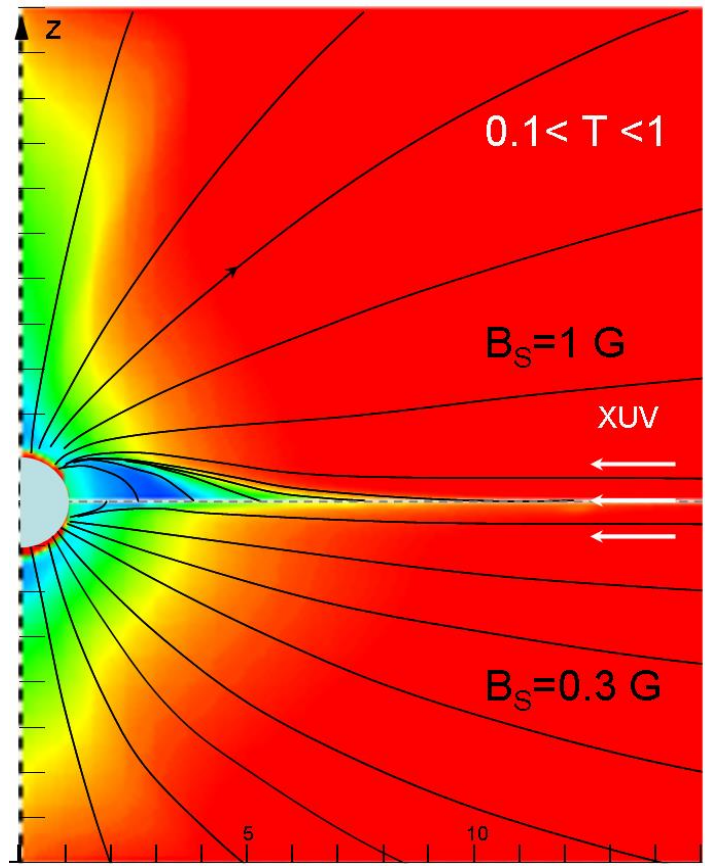

**Figure 7.** Temperature distributions calculated for a magnetized analog of HD209458b with the equatorial surface field $B_s$=0.3 G (lower panel) and $B_s$=1 G (upper panel). Magnetic field lines are shown in black.

To describe the effect of magnetic field in the simulated thermal escape of a HJ's upper atmosphere, in more details, we present in Figure 7 the spatial distribution of temperature calculated for different surface values of the planetary intrinsic magnetic dipole field $B_s$. The regions of increased temperature above the poles and in the "dead-zone" where magnetic field and tidal force suppress the plasma outflow are clearly visible.

3.2.2 *Formation and quasi-periodic dynamics of magnetodisk*

Formation of an extended and thin current sheet in the equatorial plane admits that the reconnection of reversed magnetic field lines might be triggered. It has been revealed in course of the simulations that the compression of current sheet by gas advection towards equatorial plane results in a burst-like reconnection regardless of the exact value of used conductivity or numerical dissipation. Based on the performed simulations we propose the following scenario. The stretching of open field lines by gas motion outside of the "dead-zone" tends to generate, at first stage, a thick current layer $J_\varphi \sim \sin\theta/R^2$ centred on the equatorial plane. The resulting force $J_\varphi B_R$ drives the plasma towards equator plane. This process is additionally supported by rarefication of plasma behind the dead zone cone. When the magnetic force is balanced by the thermal pressure of the compressed gas, a typical thin current sheet is formed. However, the complete force balance in the equatorial current sheet cannot be achieved because the magnetic field of the magnetodisk and thermal pressure inside it gradually decrease with the radial distance. The pressure gradient drives a flow along the equatorial plane, while gradually depleting and thinning the current sheet. When the merging of opposite field lines and conversion of magnetic energy into material jets becomes energetically preferable, an explosive reconnection starts which is driven by the pressure of the converging flow of the surrounding plasma. A part of the magnetic flux and related electric current accumulated in the magnetodisk are ejected outwards as a ring-type plasmoid, whereas the remaining magnetic flux converts into the closed field lines. Then the whole cycle repeats. The steps of this cyclic dynamics of the magnetodisk, identified in course of the performed simulation, are shown in Figure 8 as three snap-short pictures of the magnetic configuration and thermal pressure distribution at $\tau = 380, 400$ and $420$ which correspond to various dynamical phases.

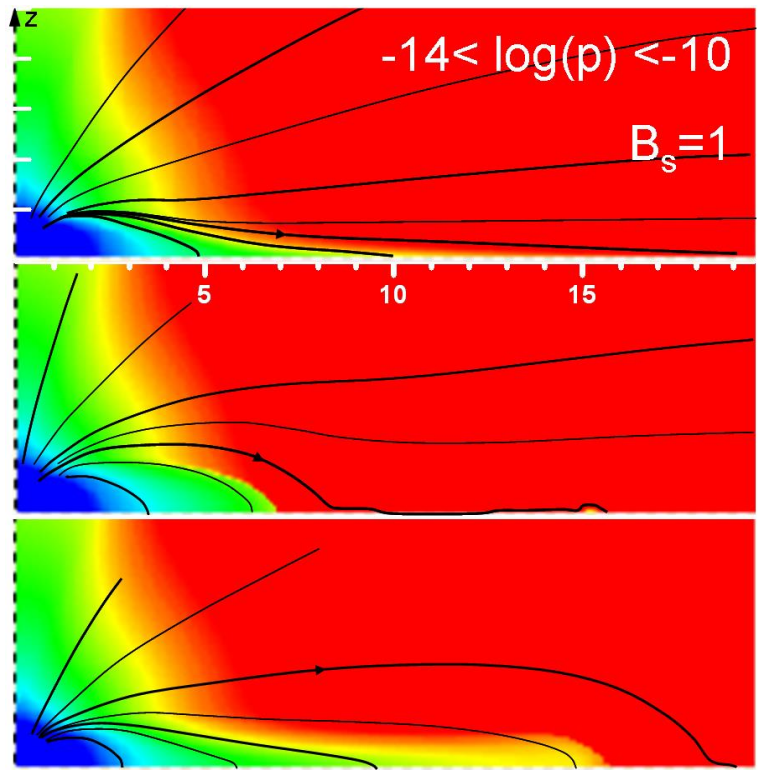

**Figure 8.** Magnetodisk cyclic dynamics (formation, reconnection, dipolarization) illustrated by the color maps of the plasma thermal pressure (in bar) distribution and by the magnetic field configuration (black lines). Upper panel: magnetodisk thinning prior to reconnection ($\tau = 380$); Middle panel: reconnection of accumulated flux and release of ring-type plasmoid ($\tau = 400$); Bottom panel: immediate post-reconnection phase, when closed field lines are being stretched again ($\tau = 420$), and the whole cycle repeats.

The reconnection may start and proceed at several radial locations in the current-carrying magnetodisk, producing the outwards propagating current filaments with the magnetic island type cross-section structure. As a result, the material outflow in equatorial region becomes strongly modulated by the cycle of magnetodisk flux accumulation and short explosive relaxation. This also modulates in time the total (instant) mass loss rate of the HJ (see in section 3.2.3). The frequency of consequent reconnection events increases with the planetary magnetic field. Indeed, the larger is the planetary field, the larger magnetic force which drives plasma towards the equatorial plane, and the faster the threshold of reconnection is reached.

Typical example of a reconnection event found in simulations is shown in Figure 9. A small region 1.25 x 1.25 centred on the reconnection point is shown at the left panel. The reconnection in this case develops at the peak of the "dead-zone". Two field lines passing close to the dead zone boundary bifurcate near the Y-point with one line going away from the planet along the manetodisk and another turning towards the planet. There is also an inflow driving the magnetic field towards the Y-point and the outflow jets typical for the reconnection. The cross field velocity towards the Y-point is a super-Alfvénic one with the Alfvén Mach number $M_A \approx 3$ and super-sonic with $M_S \approx 2$. The maximum value of z-component of the velocity is $V_z \approx 0.7$, while the maximum jet velocity towards and outwards the planet is about $V_r \approx 2.3$. The right panel of Figure 9 shows magnetic island formed by the reconnected field lines.

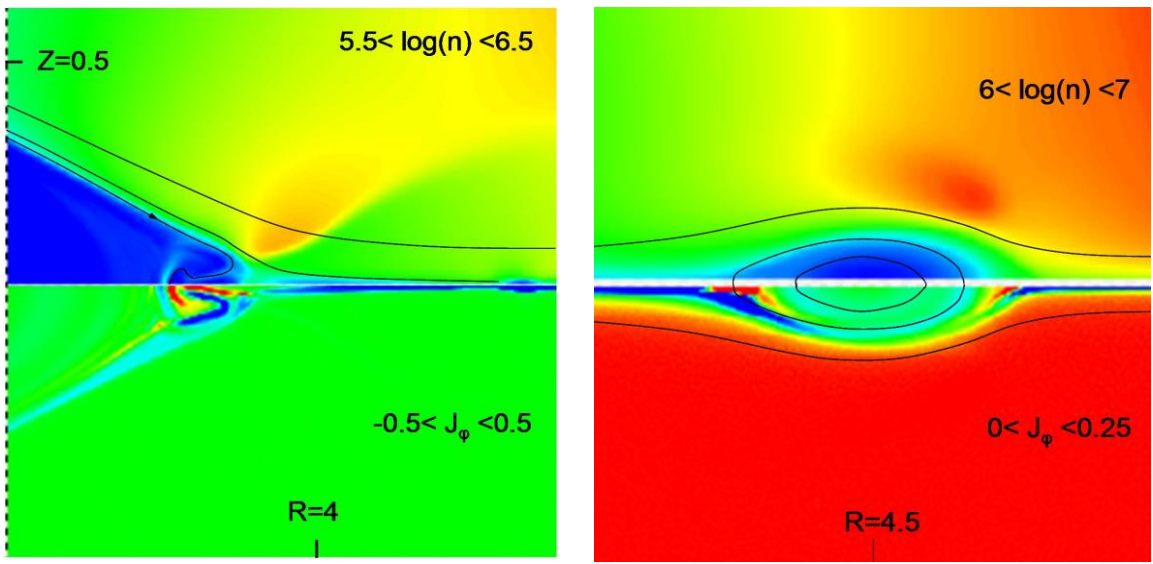

**Figure 9.** Reconnection structures illustrated by the color maps of the plasma- and electric current density distributions: Y-point (left) and magnetic island (right). By black magnetic field lines are shown.

It is interesting to note that the periodic reconnection of magnetodisk found in the present work has probably an observational analog in Jovian magnetosphere (Kronberg et al 2007). While the source of mass loading in the case of HJ is the planetary wind instead of Io, and the stretching of the field lines is also mainly due to the atmospheric material expansion and not because of the centrifugal force, the picture of the quasi-periodic dynamics of the inner magnetosphere remains rather similar. In particular, it involves the stretching and dipolarization of the magnetic field and the ejection of plasmoid in both cases.

3.2.3 *"Dead-zone" scaling and structure.*

As important parameters of the HJ's inner magnetosphere topology, relevant also in the context of potential observations, appear the size and shape of the "dead-zone" and its dependence on the planetary magnetic field value. In particular, the internal structure of the "dead-zone" is characterized by variation of plasma parameters in the direction across the equatorial plane (i.e., along the Z-axis). The graphs in Figure 10, plotted at fixed $r = 2.5\ R_p$ in case of $B_s = 1G$, indicate the presence of a rather sharp transition boundary around $Z_{tr} \approx 0.9\,R_p$ at which the thermal pressure drops while the plasma speed jumps from zero to the supersonic values. The comparison of thermal and magnetic pressures within the radial distance range $0<r<1.5\,R_p$ reveals, that their sum varies along Z-axis only by 10%. This means that the increased plasma pressure inside of the "dead-zone" is balanced by magnetic force.

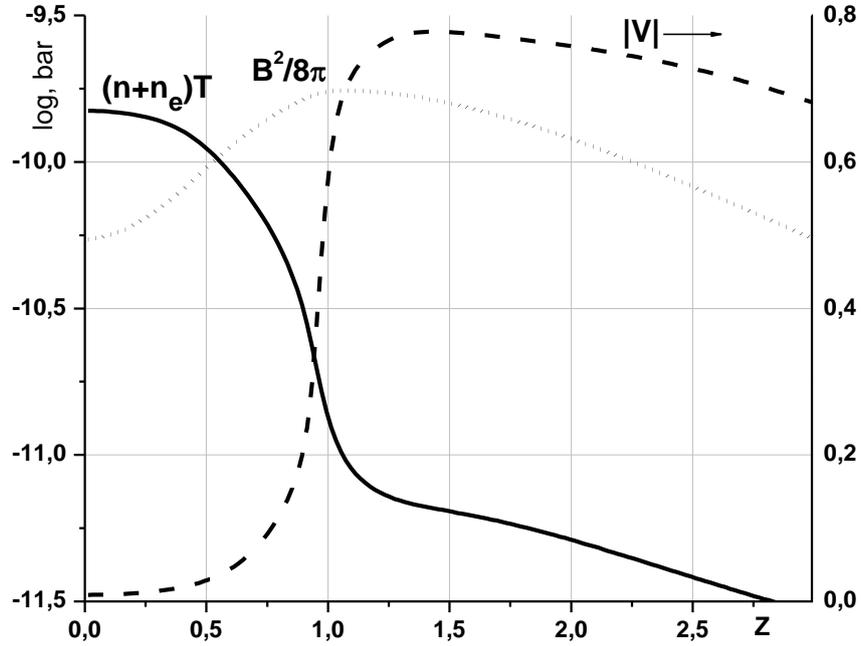

**Figure 10.** Profiles of thermal plasma pressure (solid line) and scalar magnetic pressure (dotted line) along Z-axis (scaled in $R_p$) at fixed $r=2.5\,R_p$ for the case of surface magnetic field of $B_s = 1$ G. The right axis scales the profile of plasma speed (dashed line).

The shape of the "dead-zone" could be characterized by the radial dependence of its lateral boundary position $Z_{tr}(r)$ in the direction across the equatorial plane. By means of $Z_{tr}(r)$, the cross-section $S_{DZ} = \int Z_{tr}(r)\,dr$ of the "dead-zone" can be found, as well as its maximal transverse size $Z_{tr,max}$. The length of the "dead-zone" (i.e. the size in the radial direction) is not marked by a sharp change of physical values like in the case of the lateral boundary. In that respect, the equatorial Alfvénic point, i.e. the distance $R_A$ at which plasma speed reaches the local Alfvén speed, may be taken as an indicator of the position of the "dead-zone" tip-point in the equatorial plane, or the so called "dead-point". This "dead-point" is located in close vicinity of the region where the helm-like structure of the "dead-zone" transforms to the magnetodisk. At the same time, it is important to keep in mind that the overall equilibrium of the inner quasi-stationary "dead-zone", indicated also by the comparative pressures study, is to certain extend defined by the balance between the thermal pressure of the hot plasma, confined within the "dead-zone", and the magnetic pressure.

The dependences of $R_A$, $Z_{tr,max}$ and $\sqrt{S_{DZ}}$, shown in Figure 11, reveal that the "dead-zone" size scales somewhat faster than $\sim B_s^{1/3}$ predicted for the isothermal wind and pure dipole-type planetary field case (Adams 2011). One of the reasons is that not only magnetic field changes with distance, but the thermal gas pressure does as well. In the range of radial distances (1.5 - 5) $R_p$ it decreases as $p \sim r^{-(2.5 \div 3.5)}$. Thus, the "dead-zone" scaling derived from the balance between thermal pressure and the dipole-type planetary magnetic field, is $\sim B_s^{0.6 \div 0.8}$, which is much closer to one observed in numerical simulations ($\sim B_s^{0.55 \div 0.7}$, depending on the scaling criterion taken). However, more detailed analysis shows that the radial distance to the pressure balance point $R_\beta$ where $p = B_z^2/8\pi$ is in fact about 1.5-2 times shorter than the actual length of the "dead-zone". Moreover, at distances between $R_\beta$ and $R_A$ the magnetic field decreases much faster than the dipole field because in this region the plasma flow develops which stretches field lines. Therefore it may be concluded, that the "dead-zone" consists in fact of two parts: 1) the *inner* region with stagnant plasma and 2) the *outer* region where magnetic field lines still remain closed, but the plasma flow gradually develops. The extended *outer* region exists in the "dead-zone" because the whole system does not reach a static state and undergoes periodic re-configurations. In an asymptotic steady state it should contract to a rather thin transition layer near the bifurcated magnetic field line.

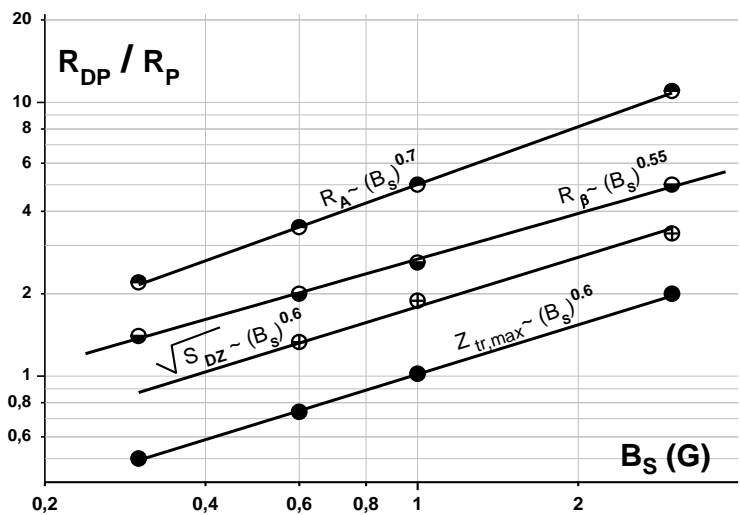

**Figure 11.** The dependences of the characteristic scales of the "dead-zone" on the value of the planetary surface magnetic dipole field $B_s$ at the equator. The values and corresponding fits are shown for the maximum width of the "dead-zone" $Z_{tr,max}$, a square root of the "dead-zone" cross-section $S_{DZ}$, an equatorial distance to the thermal and magnetic pressures balance point $R_\beta$, and that to Alfven point $R_A$. All distances are in units of the planetary radius $R_p$.

*3.2.4 Magnetically controlled mass loss of HJ's atmosphere.*

After a typical interval of about 200 characteristic time scales $R_p/\tilde{V}_c$ since the calculation start, a quasi-stationary planetary wind settles. The reconnection events strongly modulate in time the instant mass loss rate measured through the outer boundary of the simulation box. With the increase of the surface magnetic field, the quasi-periodic mass loss pulses become more pronounced in amplitude and more frequent in time. The dynamics of the mass loss rate for different values of the magnetic field is presented in Figure 12.

The time averaged dependence of the mass loss rate on planetary surface magnetic field is presented in Figure 13. As it can be seen in both Figures 12 and 13, sharp decrease of the mass loss rate takes place for the magnetic field values beyond $B_s \sim 0.3$ G, and at $B_s = 1$ G the averaged mass loss rate becomes about an order of magnitude smaller than that without the magnetic field. This effect may be explained by the increase of polar- and dead-zones areas of the suppressed material outflow with the increasing planetary magnetic field at the expense of the size of the wind region. Note that the mass loss rate in the non-magnetized case ($\sim 3\cdot 10^{10}$ g s$^{-1}$) is about two times less than that reported in our previous study (Shaikhislamov et al. 2014). This difference is because of the use there of the more idealized 1D model with a simplified description of the dynamics of atmospheric composition.

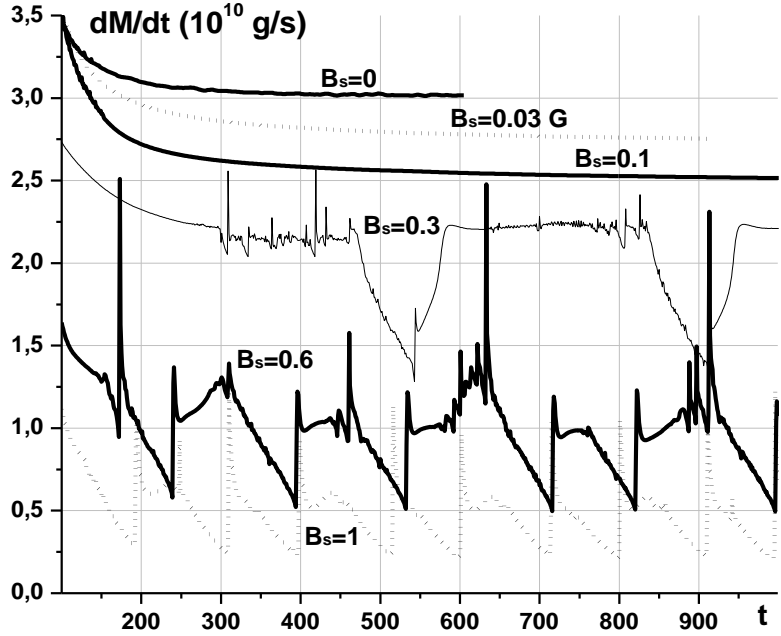

**Figure 12.** The dynamics of mass loss rate of an HD209458b analogue planet for different values of planetary surface magnetic dipole field $B_s$ at the equator. Pulsations of the mass loss rate for $B_s >0.1$ G are related with the quasi-periodic character of magnetodisk formation and evolution and the ejection of ring-type plasmoids during the reconnection events.

The accumulated flux of magnetodisk $\Phi_m(t)$ can be expressed via the magnetic flux function $F_m(r,z,t) = r\cdot A_\varphi(r,z,t)$ as follows:

$$\Phi_m(t) = \Delta\varphi \cdot r \int_o^{\Delta z} B_r(t) dz = \Delta\varphi \cdot [F_m(r,0,t) - F_m(r,\Delta z,t)], \qquad (5)$$

where $\Delta\varphi$ is azimuthal size of a sector of the magnetodisk and $\Delta z$, its half width. Due to reconnection, the flux function at the current sheet center (i.e. for z=0) varies from the maximum value $F_{m,max} = F_m(r, z=0, t_{max})$ reached at time $t_{max}$ of maximum compression of the magnetodisk, to a minimum value reached immediately after the reconnection at $t_{min}$ and equal to the flux function value at the current sheet lobes $F_{m,min} = F_m(r, z=0, t_{min}) = F_m(r, \Delta z, t_{max})$ at the moment of maximal compression $t_{max}$. Thus, by measuring of the periodic variation of the flux function at the equatorial plane outside of the dead zone, the maximal magnetic flux stored in the magnetodisk can be found as $\Phi_{max} = \Delta\varphi \cdot (F_{m,max} - F_{m,min})$. Dedicated simulation runs have shown that this value doesn't depend on the radial distance $r$. This fact reflects the conservation of the magnetic flux of the magnetodisk.

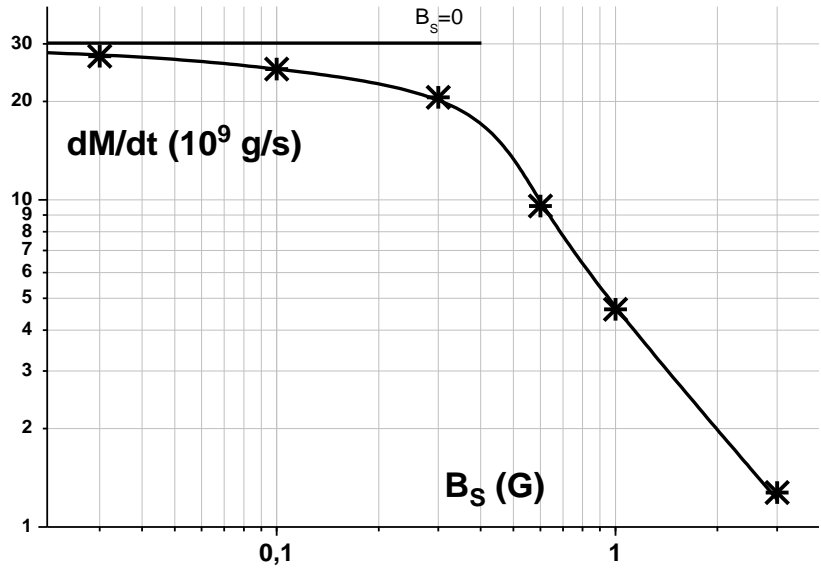

**Figure 13.** Time averaged mass loss rate of an HD209458b analog planet as a function of planetary magnetic field value $B_s$. Horizontal line indicates the mass loss rate without magnetic field.

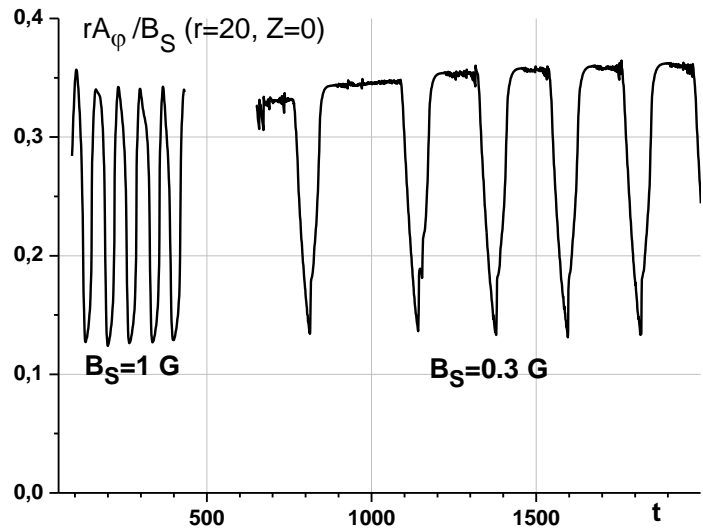

**Figure 14.** Dynamics of a normalized magnetic flux function at the center of magnetodisk current sheet $r \cdot A_\varphi(r=20, z=0, t)/(R_p^2 B_s)$ for two values of planetary magnetic field $B_s$.

The normalized flux $\Phi_{max} = \tilde{\Phi}_{max}/(R_p^2 B_s \Delta\varphi)$ was found to be equal to about 0.2 for different values of surface field (e.g., $B_s=1$ and $B_s=0.3$ G). This effect is demonstrated in Figure 14. For comparison, the initial normalized dipole field flux $\tilde{\Phi}_o = \int_o^{\pi/2}(B_r/B_s)d\theta = 2$ is an order of magnitude larger. The magnetodisk flux is composed of only a part of the field lines of the "wind-zone", which itself is restricted between the equatorial dead zone and suppressed polar zone.

*3.2.5 The pressure portrait of the magnetized planetary wind*

To imagine how the planetary wind and magnetodisk might interact with the incoming stellar wind, we plot in Figure 15 the profiles of different pressures, transverse to the wind (i.e. along $Z$ axis) at a fixed radial distance $r=10R_p$ from the planet. Scalar magnetic pressure $B^2/8\pi$ and the total pressure $B^2/8\pi + nm_p V^2 +(n+n_i)T$, where $m_p$ stays for the proton mass, are shown for $B_s=0.3$ G and $B_s=1$ G. For comparison, the wind pressure in the non-magnetized case is also shown in Figure 15, calculated for two geometric approximations, i.e. *sector* and *quasi-axisymmetric* ones. Altogether, the plots in Figure 15 show that at the distance of 10 planetary radii the escaping plasma wind of an HD209458b analog planet exerts the pressure which exceeds the typical stellar wind pressure of a solar-type star $\approx 3 \cdot 10^{-6}$ μbar at the orbital distance of ~0,05 AU (see e.g. Fig.2 in Kislyakova et al. 2014, Johnstone et al. 2015). This means that the magnetospheric obstacle of such HJ may extend even further from the planet than the considered distance of $10R_p$.

As it can be deduced from Figure 15, at considered radial distance from the planet the pressure of the magnetic field generated by gas outflow and associated with the magnetodisk is much larger than that of the initial planetary dipole field. Moreover, around the equatorial plane the resulting magnetic field is much larger than the field $B_R = 2B_s z/r^3$, expected at pure radial stretching. Therefore, magnetodisk is a dominant magnetic structure, which determines the magnetic pressure contribution in the total pressure.

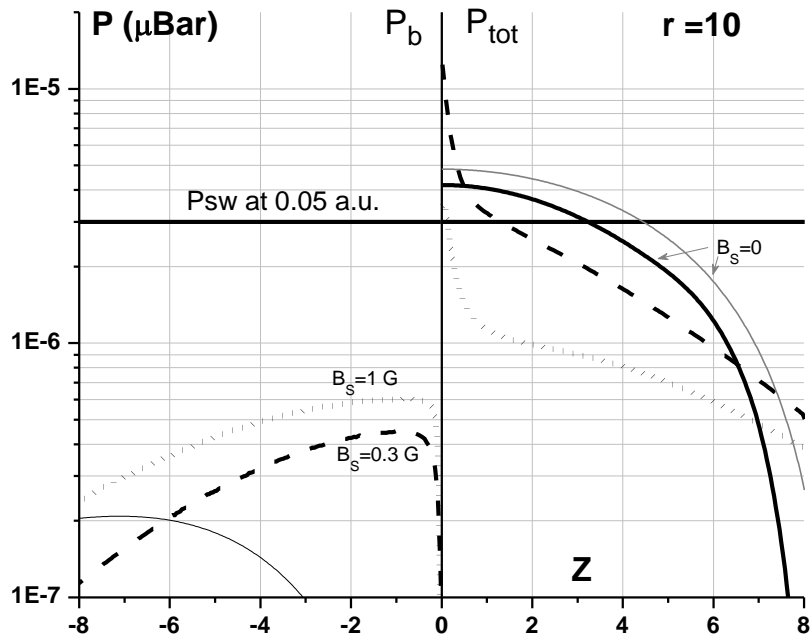

**Figure 15.** Pressure profiles along $Z$ axis (transverse to orbital plane) scaled in $R_p$ at a radial distance $r=10R_p$ towards the star along the planet-star line for the cases of planetary magnetic field $B_s=0.3$ G (dashed) and $B_s=1$ G (dotted). Negative $Z$ quadrant is used to show the scalar magnetic pressure, whereas positive $Z$ quadrant presents the total pressure (i.e., the sum of magnetic, ram, and thermal pressures). Thick and thin solid lines in the positive $Z$ quadrant show the total pressure in the non-magnetized case, calculated in *sector* and *quasi-axisymmetric* geometry approximations, respectively. Thin solid line in the negative $Z$ quadrant indicates the scalar pressure of a pure radial magnetic field in the case of $B_s=0.3$ G.

At the considered distances from the planet the ram pressure of escaping plasma wind is much larger than the thermal pressure. This is mostly due to tidal force which increases the outflow velocity by several times. It is also important to note, that the planetary magnetic field in the considered case decreases the total planetary wind ram pressure because of the contraction of the wind zone and re-structuring of the escaping plasma flow. However, due to the additional ram pressure of the outflowing gas, the total pressure inside magnetodisk is several times larger than that outside (see in Figure 15). Outside of the magnetodisk the total pressure decreases sharply. In particular, for $B_s = 1 G$ it is about 4 times smaller than that in the non-magnetized case. Such a pressure peak in the area of magnetodisk may affect the magnetosphere structure near the equatorial sub-stellar point.

## 4. Discussion and conclusions

The whole complex of the performed self-consistent MHD simulations reported in the present paper, confirms the previous qualitative guess that the escape of upper atmospheric material of a HJ heated and ionized by the stellar radiation in the presence of planetary magnetic dipole field is of essential importance for the formation and shaping of the planetary magnetosphere. At the same time, the magnetic field itself appears an important factor which influences the motion of the expanding upper atmospheric material of a HJ. Understanding and quantitative self-consistent treatment of the planetary mass loss and nearby plasma environment, as well as the identification of potentially observable features of the magnetized close-orbit giant exoplanets, is an important task. It is related with the investigation of internal structure of the planetary magnetosphere, formed under the conditions of the outflowing upper atmospheric gas heated and ionized by stellar radiation. According to the theoretical semi-qualitative treatments (Mestel 1968, Adams 2011), this structure is to a significant degree defined by the balance between the magnetic and gas pressures, which regulates the size of the equatorial "dead-zone" and "wind-zone". Our modelling for the first time enables an insight of the internal structure of these crucial regions and reveals the quantitative self-consistent description of the force balance and material dynamics there.

While HD and MHD codes are fairly well developed and available, their use for simulation of expanding planetary plasma winds on exoplanets is a relatively novel area. A special challenge in that respect consists in the necessity to evaluate and then to include in the modelling the major underlying physics and most important factors which have been explored in the last years (Yelle 2004, Tian et al. 2005, García Muñoz 2007, Erkaev et al. 2005, 2007, Penz et al. 2008, Murray-Clay et al. 2009, Guo 2011, 2013, Koskinen et al. 2010, 2013, Owen & Jackson 2012, Adams 2011, Trammell et al. 2011,2014, Shaikhislamov et al. 2014, Owen and Adams 2014). It was sown that the mass loss rate is very much affected by the EUV radiation spectra (mono-energetic or solar based), the star class and age, EUV absorption (thin layer or distributed), heating efficiency, hydrogen photo-chemistry (including ionization and recombination), infrared $H_3^+$ cooling, cooling due to Lyα emission, gravitational effects (i.e. tidal force), inner boundary conditions (i.e., fixed pressure value at conventional planetary surface), etc. At the same time, most of these factors were treated in 1D geometry approximation, and their spatial non-homogeneity, for example that of heating, has been addressed so far only qualitatively (Tian et al. 2005).

2D calculations reported in the present paper show that one-side illumination and radiative heating of a tidally-locked exoplanet leads to development of a so strong zonal wind, that the absorbed radiation energy is effectively redistributed over the whole planetary upper atmosphere resulting even in a higher temperature and stronger plasma outflow at the night side. The performed study also confirmed the importance of an appropriate account of the hydrogen chemistry at the upper atmospheric layer adjacent to inner boundary of the simulation domain where the most energetic photons are absorbed as that affects the mass loss rate (reduction by approximately a factor of two) and the whole numerical solution. That may be even more crucial for exoplanets at larger orbital distances, than those typical for HJs, as the $H_3^+$ cooling area tends to increase at lower atmospheric temperatures (Chadney et al. 2015). At the same time, the rates of the most important reactions included in our model were often extrapolated from the temperature range of 300 K at which they were measured to the higher temperature conditions realized at HJs. Therefore, these, as well as the IR emission function of $H_3^+$ ion are the subject for further investigation and revaluation with the development of more complex chemical models.

In the previous attempts to incorporate planetary magnetic field a 1D solution for the hydrodynamic escape has been modified to take into account the specific geometry of magnetic field lines (Adams 2011, Trammell et al. 2011). Calculation of the size of the "dead-zone" and constructing of the solution for the "wind-zone" made it possible to estimate the overall mass loss in dependence on the planetary magnetic field strength. Further elaboration of this approach lead to application of 2D MHD codes to simulate the escaping plasma flow in a dipole magnetic field and to model the formation of the "dead-" and "wind-" zones (Owen and Adams 2014, Trammell et al. 2014). However, the atmospheric thermal expansion itself, initiated at the inner atmosphere regions, was based in these studies on rather simplified and different from paper to paper assumptions of either a mono-energetic EUV flux, or isothermal empirically estimated gas temperature and ionization. These studies neglected also the $H_3^+$ infra-red cooling and were done for the model parameter range at which the solution depends on the pressure at the inner boundary of the simulation domain. These effects have been specially investigated in Shaikhislamov et al. (2014). As a result of an incomplete and non-self-consistent modelling, the estimations obtained by different authors for the magnetic field strength, at which the outflowing plasma wind on an HD209458b analog planet is significantly suppressed, vary by more than an order of magnitude, from less than 0.3 G (Owen and Adams 2014) to more than 3 G (Trammell et al. 2014). In particular, Trammell et al. (2014), who used the isothermal approximation, obtains for a relatively small magnetic field the mass loss rate which is about an order of magnitude higher than that obtained in the present paper. Partly the difference in the estimated magnetic suppression of the mass loss is also connected with the different planetary parameters considered by different authors. For example, Owen and Adams (2014) made the calculations for an exoplanet with the Jovian parameters (i.e., $R_p=R_J$, $M_p=M_J$), for which the gravitational potential well parameter, $M_p/R_p$, is twice larger than that considered in other works. As the mass loss depends exponentially on this parameter (Owen and Adams 2014), the conditions for the material outflow have been somewhat different. Moreover, the tidal effects produced by a close parent star strongly influence the planetary wind and the related mass loss process as well (Trammell et al. 2014, and this paper). In our model, which is free of non-self-consistently prescribed parameters (e.g., ionization degree, simplified radiation spectrum, inner boundary pressure, isothermality, etc.) and empirical substitutions (e.g., magnetic field topology, radiative heating proxy, etc.), but still built with some geometry related approximations, while including most of the processes affecting the planetary wind (e.g., hydrogen chemistry, realistic XUV spectrum, tidal effects), the mass loss is found to be strongly suppressed at the magnetic field values (on equatorial surface) beyond ~1 G (see in Figure 13).

Our modelling enabled for the first time to explore the scaling, internal structure, and force balance in the "dead-" and "wind-" zones which appear crucial elements of the HJ's inner magnetosphere driven by the expanding planetary plasma wind, predicted so far only with a semi-qualitative physical reasoning. In particular, it has been shown that the equilibrium of the "dead-zone" and its transverse (i.e., cross-equatorial) size are defined mainly by the balance between the magnetic pressure and thermal pressure of the confined stagnated plasma. At the same time, the length, i.e. extension of the "dead-zone" in the equatorial plane is related also with its specific non-stationary *outer* region, where the magnetodisk flow forms, so that the distances of Alfvénic point r = $R_A$ and unit plasma beta $R_\beta$ specify this size just to a certain extent. The non-stationary *outer* region of the "dead-zone" is essentially linked with the magnetodisk and related with its formation and evolution.

To understand the effect of magnetic field on the planetary atmospheric plasma escape, let's estimate analytically the value, at which the field affects the planetary mass loss. Similar calculations, made in previous works (e.g., Trammell et al. 2014, Owen and Adams 2014), were based on the assumption of a balance between the magnetic pressure of the planetary dipole field at the equator and the ram pressure of the escaping planetary wind. With such an approach, the ram pressure can be expressed via the integral mass loss $dM/dt = 4\pi r^2 \rho V$ obtained by spherical averaging of the continuity equation, so that $P_{ram}/P_B = (2VdM/dt)/(r^2 B^2)$. If the magnetic pressure is larger than the ram pressure, then it can be assumed that the flow is fully controlled by the magnetic field. An estimate gives that for the mass loss rate of $3 \cdot 10^{10}\,g/s$, the material escape speed of $10^6\,cm/s$, and the planet size of $1.4 \cdot 10^{10}\,cm$, the critical surface field $B_s$ which balances the ram pressure at a distance of $2R_p$ is equal to only ~0.07 G. This is much smaller than the value of 0.3 G that followed from the numerical simulations.

However, such an estimate supposes a well pronounced and relatively fast motion of the material, whereas the matter of influence of the planetary magnetic dipole field on the mass loss consists in the confinement of a large portion of plasma inside the "dead-zone". In that respect the balance between the ram and magnetic pressures does not answer the question regarding the amount of the escaping material. It is the "dead-zone" which locks a part of the planetary wind plasma and prevents its escape, reducing therefore the overall mass loss. The size of the stationary inner part of "dead-zone" is determined by the balance between the magnetic pressure and the thermal pressure of the stagnated and practically not-moving plasma (instead of the ram pressure). Only in the sufficiently remote regions of a weaker magnetic field in the outer "dead-zone" the hot plasma deforms the field lines and breaks through them into the magnetodisk and the restricted "wind-zone". For a larger planetary magnetic dipole field the extent of the "dead-zone" gets also bigger at the expense of the "wind-zone". For a rough estimate one may assume that if the size of the "dead-zone" reaches that of the planet (e.g., the "dead-zone" extends up to r~$2R_p$), a significant amount of plasma appears to be locked inside the "dead-zone", and the integral mass loss becomes to be distinctly reduced.

The typical temperature profile of a HJ (see in Figures 4, 5, and 7) has the temperature maximum of about $10^4$ K at the distance of ~(1.3-2)$R_p$ Due to that, the thermal pressure behaves non-exponentially and in the range of r = (1.5-5) $R_p$ it can be empirically approximated by a cubic function: $p \approx 10^{-10}(2R_p/r)^3\,bar$. The balance between thermal and magnetic dipole field pressures gives an estimate to the dead-point distance: $(R_\beta/R_p)^3 \approx 50(B_S/1G)^2$. As it was discussed in the section 3.2.3, this estimate corresponds in fact to the inner region of the dead zone. Thus, the condition of a large "dead-zone", i.e. $R_\beta = 2R_p$ is realized at equatorial surface field of 0.4 G. This is much closer to the equatorial surface magnetic field value (~0.3 G) followed from our numerical simulations, at which the planetary magnetic field starts to influence the mass loss (see in Figure 13).

Among the advantages of the performed self-consistent 2D modelling there is also the detailed description of the spatial distribution of the expanding planetary wind and its essentially sectorial structure. In particular, besides of the "dead-zone", it appears that the plasma flow in polar regions is strongly decelerated as well. The deceleration along the rotation axis is caused by the stellar gravity regardless of the presence of a background magnetic dipole field. This effect was described first in 1D calculation in Trammell et al. (2011) and later verified with 2D simulation in Trammell et al. (2014). The action of the tidal force in the case of HD209458b analog planet results in confinement of the whole planetary plasma wind within a kind of a tube aligned with the planet-star line and formed by two surfaces placed symmetrically on both sides of the ecliptic at the model predicted distance of about $16R_p \approx 4R_{L1}$. However, despite of strong influence of the tidal force on the structure and velocity profile of the escaping material flow, the overall mass loss rate increases insignificantly with the account of the stellar gravitation effects.

Inclusion of the background planetary magnetic dipole field into consideration results in additional suppression of the escaping material flow in the polar regions. While in an isothermal case (Adams 2011) the faster polar divergence of the plasma flow guided along the magnetic field lines enables conditions for the super-sonic flow regime, the effect of adiabatic cooling of the diverging flow taken into account in the self-consistent non-isothermal model, decreases the flow acceleration and finally prevents achievement of the sonic point at all.

The results of performed MHD modelling are consistent with the theoretical prediction (Khodachenko et al. 2012) and laboratory experiment (Antonov et al. 2013) regarding the fact that the flow of a HJ's escaping upper atmospheric material, interacting with the planetary magnetic dipole field generates magnetodisk located beyond the "dead-zone". Based on the numerical simulations, the process of magnetodisk formation, driven by the thermal and ram pressures of the heated by stellar XUV and expanding plasma which stretches the planetary magnetic dipole field lines, could be viewed in deeper details and understood better. To a certain extent, this process is analogous to the processes of the Jovian magnetodisk, as well as the heliospheric current sheet formation, though the primary energy source in the case of an exoplanetary magnetosphere, is the absorbed stellar XUV radiation converted into thermal gas pressure, rather than rotation (as for Jupiter and heliosphere) or kinetic (as for heliosphere) energy.

The essentially novel feature discovered in the present work, and not mentioned so far in the previous theoretical qualitative and numerical studies, is the quasi-periodic dynamical character of the HJ's magnetodisk evolution with the repeating phases of accumulation of the magnetic flux and its following relaxation via fast reconnection, accompanied by release of a ring-type plasmoid. The frequency of the reconnection events (i.e. the magnetodisk evolution cycle) increases with the increasing planetary intrinsic magnetic moment. The portion of the periodically accumulated and reconnected magnetic flux was found to constitute about 10% of the total magnetic dipole flux through the planet surface. Such ejections of plasmoids and particle acceleration accompanying the explosive reconnection events which should happen with a regular periodicity might be of interest in the context of possible observations and remote probing of the HJs' environment. Another potentially observable phenomenon quantified in our present study is the "dead-zone" which is formed below the magnetodisk and which contains a relatively hot ~$10^4$ K and dense >$10^6$ cm$^{-3}$ stagnant plasma, co-moving with the planet. The "dead-zone", therefore may be a reason for the observed nowadays and widely discussed effects of early ingress in the spectra of exoplanetary transit observations (Fossati et al. 2010, Llama et al. 2011, Bisikalo et al. 2013).

All the models proposed so far for the study of thermal escape of HJs' upper atmospheric material and related with that planetary *thermal* mass loss, including the present paper, need further upgrade to include the next crucial element of the phenomenon's picture – the stellar wind. This would enable to quantify the issues regarding the size and shape of the planetary magnetospheric obstacle and the character of plasma motion inside and outside of it. This information in its turn is important for the description of the interaction of the escaping planetary atmospheric particles with the stellar wind flow and a pick-up process, which constitutes the planetary *non-thermal* mass loss mechanism. Our estimations show, that either the thermal, or ram pressure of the escaping plasma wind of an HD209458b analog planet exceed the stellar wind ram pressure up to the distances of (5-10) $R_p$ (e.g., Figure 15). Beyond these heights, the interaction of a tidally locked non-magnetized HJ with the stellar wind can be modelled in the *quasi-axisymmetric* geometry approximation, with the asymmetry of the centrifugal force in terminator plane somewhat compensated by the more or less axisymmetric pressure of the stellar wind. By this, the total mass loss rate should not be significantly modified due to the hydrodynamic interaction with the stellar plasma flow, since the already escaped (thermally) planetary atmospheric material directed towards the star will be then just picked-up by the stellar wind and blown away from the planet and from the star, or redirected into the tail.

Due to magnetic fields frozen into stellar wind the interpenetration boundary of planetary and stellar protons should be very thin compared to the magnetosphere size, characterized by the magnetopause stand-off distance. Planetary atmosphere atoms, however, should penetrate freely into the stellar wind and after the charge exchange produce energetic neutral atoms (ENAs) which could be seen at the wings of Lyα absorption profile (Ben-Jaffel 2007, Tremblin & Chiang 2013). Self-consistent calculation of such an observable effect is of great interest and is planned as a next step of our studies. That would enable more precise characterization of the planetary ENA coronas and related with that magnetospheric obstacle scales and shapes (Holmström et al. 2008, Ekenbäck et al. 2010, Kislyakova et al. 2014).

Finally, we would like to note that on magnetized exoplanets the stellar wind might significantly change the character and shape of magnetodisk, especially in the case of a relatively small magnetosphere. It is very likely that the magnitodisk-type structure will be stretched towards the tail where also the whole planetary wind will be channeled. The "night-side" magnetodisk might still exhibit periodic reconnection. However, the reconnecting flux will be supplied by the planetary magnetic dipole field lines torn off by the planetary wind independently on the interplanetary magnetic field transported by the stellar wind. Simulation of such problem requires a 3D model which remains a challenging task. Therefore, the study of hydrodynamic escape of a magnetized close-orbit giant exoplanet with a geometrically simplified model and idealized outer boundary conditions presented in this paper, is a necessary step which enables the estimation of the planetary mass loss rate and reveals several new effects that must be taken into consideration in more complex models.

**Acknowledgements**

This work was supported by grant №14-29-06036 of the Russian Fund of Basic Research, RAS presidium program N9, RAS SB research program (project II.10.1.4, 01201374303), as well as by the projects S11606-N16 and S11607-N16 of the Austrian Science Foundation (FWF). MLK also acknowledges the FWF projects P25587-N27, P25640-N27 and the The Leverhulme Trust grant IN-2014-016. The crucial for the present study parallel computations have been performed at the Supercomputing Center of the Lomonosov Moscow State University and at the Siberian Super-Computing Center (SSCC). The authors are also thankful to EU FP7 project IMPEx for providing collaborative environment for research and communication.

**Figure Captions**

**Figure 1.**
Two cases of geometry approximation applied for calculation of planetary wind and related mass loss: *(a) quasi-axisymmetric* approximation, suitable for a non-magnetized planet or a planet with magnetic dipole oriented along the planet-star line; and *(b) sector* approximation suitable for a magnetized planet.

**Figure 2.**
Color maps of temperature T, ionization degree $X_{ion} = n_i/(n_i + n_a)$, radial (spherical) Vr and zonal Vθ velocities, calculated in the *quasi-axisymmetric* geometry approximation for a non-magnetized analog of HD209458b in the cases with 1) zero tidal force, i.e. $R_{L1} = \infty$ (top panels), and 2) a realistic tidal force, i.e. $R_{L1} = 4.53 \cdot R_p$ (bottom panels). Variation ranges of the simulated physical parameters are indicated in the top of the corresponding panels. Adopted color scale between the normalized minimal and maximal values of a simulated physical parameter is presented in the upper right panel. All values outside the considered variation range of each parameter, i.e. below or above the corresponding minimal and maximal values, are plotted as uniformly red or blue, respectively. The black lines show the material flow streamlines.

**Figure 3.**
Spatial distribution profiles of radial velocity (top panel) and plasma density $n = n_H + n_i$ (bottom panel), calculated for a non-magnetized analog of HD209458b in *sector* approximation geometry with Z-axis directed perpendicular to the ecliptic and the planet-star line. Black lines show the streamlines of the planetary wind flow.

**Figure 4.**
Profiles of the planetary wind plasma velocity (solid), temperature (dashed) and density $n = n_H + n_i$ (dotted) in the Z-axis direction, i.e. perpendicular to the planet-star line (left panel) and in the direction towards the star (right panel), scaled in $R_p$. Thick and thin lines correspond to the calculations in *sector* and *quasi-axisymmetric* geometry approximations, respectively. Arrows point to the graph axis which shows scaling of a particular parameter.

**Figure 5.**
Spatial variation of the EUV absorption rate (thick solid line), advection and adiabatic cooling rate (dashed line), as well as radiative cooling rate due to infrared (i.e. $H_3^+$) and Lyα emission (dotted lines) along the planet-star line (scaled in $R_p$), calculated for a non-magnetized analog of HD209458b within *sector* geometry approximation. Gray lines show profiles of temperature and relative content of molecular hydrogen along the planet-star line (values are given at the right axis).

**Figure 6.**
Color maps of the radial velocity and particle density distributions calculated for a magnetized analog of HD209458b with equatorial surface field $B_s$=0.3 G (left panel) and $B_s$=1 G (right panel). Magnetic field lines are shown in black. For comparison the dipole field lines are plotted by dashed gray lines.

**Figure 7.**
Temperature distributions calculated for a magnetized analog of HD209458b with the equatorial surface field $B_s$=0.3 G (lower panel) and $B_s$=1 G (upper panel). Magnetic field lines are shown in black.

**Figure 8.**
Magnetodisk cyclic dynamics (formation, reconnection, dipolarization) illustrated by the color maps of the plasma thermal pressure (in bar) distribution and by the magnetic field configuration (black lines). Upper panel: magnetodisk thinning prior to reconnection ($\tau = 380$); Middle panel: reconnection of accumulated flux and release of ring-type plasmoid ($\tau = 400$); Bottom panel: immediate post-reconnection phase, when closed field lines are being stretched again ($\tau = 420$), and the whole cycle repeats.

**Figure 9.**
Reconnection structures illustrated by the color maps of the plasma- and electric current density distributions: Y-point (left) and magnetic island (right). By black magnetic field lines are shown.

**Figure 10.**
Profiles of thermal plasma pressure (solid line) and scalar magnetic pressure (dotted line) along Z-axis (scaled in $R_p$) at fixed $r$=2.5 $R_p$ for the case of surface magnetic field of $B_s = 1$ G. The right axis scales the profile of plasma speed (dashed line).

**Figure 11.**
The dependences of the characteristic scales of the "dead-zone" on the value of the planetary surface magnetic dipole field $B_s$ at the equator. The values and corresponding fits are shown for the maximum width of the "dead-zone" $Z_{tr,max}$, a square root of the "dead-zone" cross-section $S_{DZ}$, an equatorial distance to the thermal and magnetic pressures balance point $R_\beta$, and that to Alfven point $R_A$. All distances are in units of the planetary radius $R_p$.

**Figure 12.**
The dynamics of mass loss rate of an HD209458b analog planet for different values of planetary surface magnetic dipole field $B_s$ at the equator. Pulsations of the mass loss rate for $B_s > 0.1$ G are related with the quasi-periodic character of magnetodisk formation and evolution and the ejection of ring-type plasmoids during the reconnection events.

**Figure 13.**
Time averaged mass loss rate of an HD209458b analog planet as a function of planetary magnetic field value $B_s$. Horizontal line indicates the mass loss rate without magnetic field.

**Figure 14.**
Dynamics of a normalized magnetic flux function at the center of magnetodisk current sheet $r \cdot A_\varphi (r=20, z=0, t) / (R_p^2 B_s)$ for two values of planetary magnetic field $B_s$.

**Figure 15.**
Pressure profiles along $Z$ axis (transverse to orbital plane) scaled in $R_p$ at a radial distance $r=10R_p$ towards the star along the planet-star line for the cases of planetary magnetic field $B_s=0.3$ G (dashed) and $B_s=1$ G (dotted). Negative $Z$ quadrant is used to show the scalar magnetic pressure, whereas positive $Z$ quadrant presents the total pressure (i.e., the sum of magnetic, ram, and thermal pressures). Thick and thin solid lines in the positive $Z$ quadrant show the total pressure in the non-magnetized case, calculated in *sector* and *quasi-axisymmetric* geometry approximations, respectively. Thin solid line in the negative $Z$ quadrant indicates the scalar pressure of a pure radial magnetic field in the case of $B_s=0.3$ G.